\documentclass[preprint,journal]{vgtc}            

\onlineid{1770}

\preprinttext{To appear in IEEE Transactions on Visualization and Computer Graphics.}

\vgtcpapertype{Analytics \& Decisions}

\title{Sel3DCraft: Interactive Visual Prompts for User-Friendly Text-to-3D Generation}

\author{%
  \authororcid{Nan Xiang\textsuperscript{\dag}}{0009-0008-8478-2391},
  \authororcid{Tianyi Liang\textsuperscript{\dag}}{0000-0001-8372-8379},
  \authororcid{Haiwen Huang}{0009-0005-9443-3855},
  \authororcid{Shiqi Jiang}{0000-0002-8175-874X},
  Hao Huang,
  \authororcid{Yifei Huang}{0000-0002-3077-0175},
  \authororcid{Liangyu Chen}{0009-0005-0243-3613},\texorpdfstring{\\}{}
  \authororcid{Changbo Wang}{0000-0001-8940-6418}, and
  \authororcid{Chenhui Li}{0000-0001-9835-2650},
}

\authorfooter{
  \item N. Xiang\textsuperscript{\dag}, T. Liang\textsuperscript{\dag}, H. Huang, S. Jiang, H. Huang, Y. Huang, L. Chen, C.~Wang, and C.~Li are with East China Normal University. \textsuperscript{\dag}These authors contributed equally. Chenhui Li and Liangyu Chen are corresponding authors. Email: \{51275902056, tyliang, hwhuang, 52265901032, 10215102453\}@stu.ecnu.edu.cn, yifeihuang17@gmail.com, lychen@sei.ecnu.edu.cn, \{cbwang,chli\}@cs.ecnu.edu.cn.
}

\abstract{
Text-to-3D (T23D) generation has transformed digital content creation, yet remains bottlenecked by blind trial-and-error prompting processes that yield unpredictable results. While visual prompt engineering has advanced in text-to-image domains, its application to 3D generation presents unique challenges requiring multi-view consistency evaluation and spatial understanding. We present \textit{Sel3DCraft}, a visual prompt engineering system for T23D that transforms unstructured exploration into a guided visual process. Our approach introduces three key innovations: a dual-branch structure combining retrieval and generation for diverse candidate exploration; a multi-view hybrid scoring approach that leverages MLLMs with innovative high-level metrics to assess 3D models with human-expert consistency; and a prompt-driven visual analytics suite that enables intuitive defect identification and refinement. Extensive testing and user studies demonstrate that \textit{Sel3DCraft} surpasses other T23D systems in supporting creativity for designers.
}


\teaser{
  \centering
  \includegraphics[width=\linewidth, alt={A view of a city with buildings peeking out of the clouds.}]{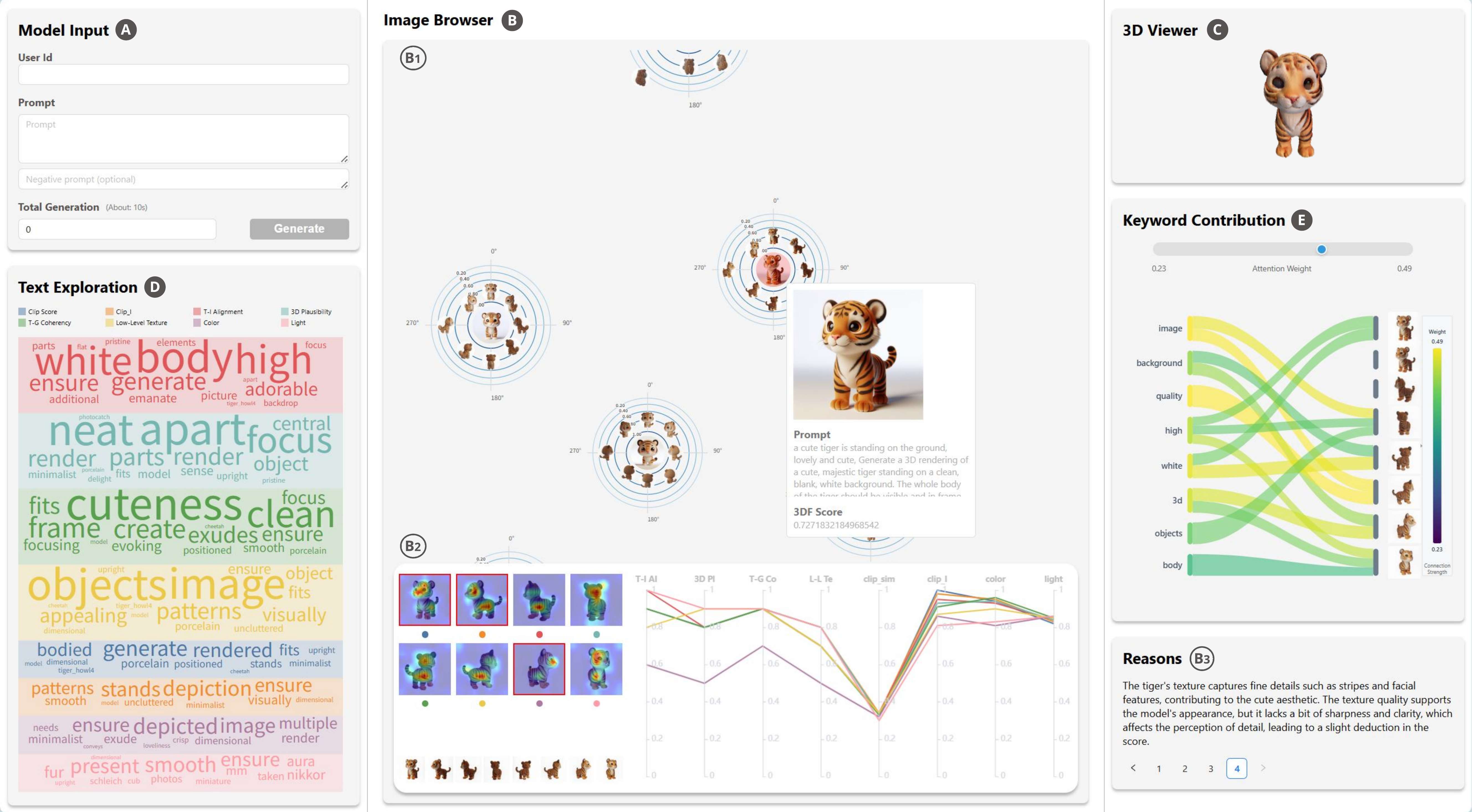}
  \caption{
  The user interface of \textit{Sel3DCraft} includes the Model Input View (A); the Image Browser View (B), which covers the generated and retrieved images (B1), the multi-view hybrid-level scoring function (B2), and the scoring rationale (B3); the 3D Viewer View (C); the Text Exploration View (D); and the Keyword Contribution View (E).
  }
  \label{fig:teaser}
}

\renewcommand{\manuscriptnotetxt}{}

\nocopyrightspace

\graphicspath{{figs/}{figures/}{pictures/}{images/}{./}} 

\usepackage{tabu}                      
\usepackage{booktabs}                  
\usepackage{lipsum}                    
\usepackage{mwe}                       

\usepackage{mathptmx}                  

\usepackage{marginnote}           
\usepackage{adjustbox}            
\usepackage{amsmath}              
\usepackage{amssymb}              

\newif\ifrevise
\revisefalse    

\global\marginparsep=3pt

\newcommand{\sidecomment}[1]{%
  \ifrevise
    \marginnote{%
      \textcolor{red}{%
        \adjustbox{minipage=0.7\marginparwidth,fbox}{%
          \scriptsize#1%
        }%
      }%
    }%
  \fi
}

\newcommand{\revision}[1]{%
  \ifrevise
    {\hypersetup{allcolors=red}\color{red}#1}%
  \else
    #1%
  \fi
}

\newcommand{\revisionbox}[1]{%
  \ifrevise
    \setlength{\fboxrule}{0.5pt}
    \setlength{\fboxsep}{1pt}
    \fcolorbox{red}{white}{#1}%
  \else
    #1%
  \fi
}

\begin{document}


\firstsection{Introduction}

\maketitle
Text-to-3D (T23D) generation, which transforms textual prompts directly into 3D models, has revolutionized digital content creation across industries such as virtual reality, gaming, film, and animation. Despite reducing model creation time, a critical gap remains between user expectations and generated output quality.

\begin{figure}
    \centering
    \includegraphics[width=1.0\linewidth]{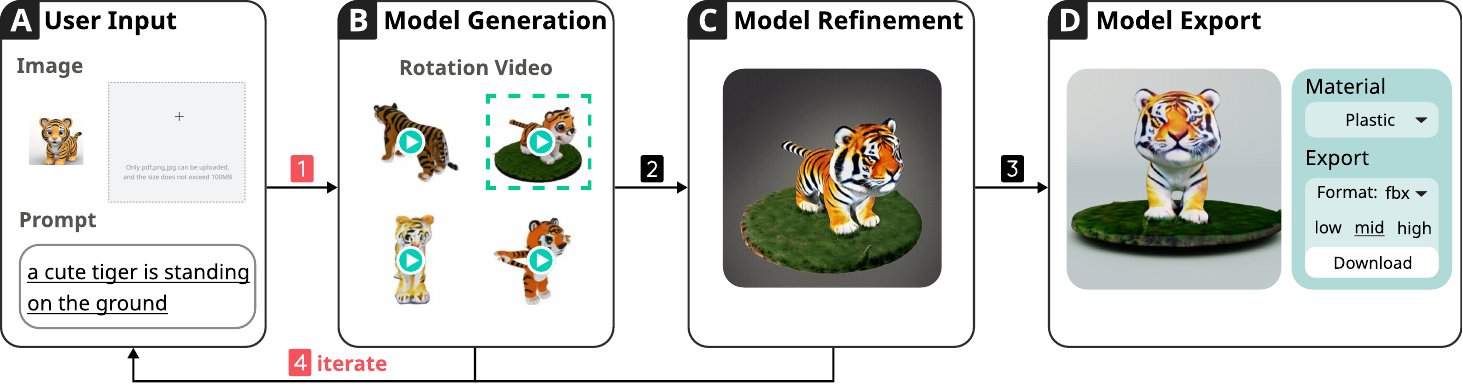}
    \caption{The workflow of text-to-3D generation tools involves four steps: (1) users input text or images and receive four rotating previews of coarse models; (2) a single model is selected for time-intensive refinement; (3) the final model is exported; and (4) if unsatisfied, users can iterate on prompts during generation or refinement. Our system enhances the first and fourth steps, improving efficiency in prompt engineering and model selection.}
    \label{fig:mainprocess}
            \vspace{-14pt}
\end{figure}

\revision{Current T23D tools operate in an end-to-end ``black-box'' manner (see \cref{fig:mainprocess}). They offer limited control over the generation process and frequently produce models that fail to meet expectations~\cite{tang2023dreamgaussian,liu2023zero,hong2023lrm}. The absence of effective prompt recommendation mechanisms forces users into costly trial-and-error cycles. This disrupts creative workflows and limits the technology's practical utility for professional designers.}

\sidecomment{R2.Q2\\SR.Q2}\revision{Recent advances in prompt engineering have shown promise in text-to-image domains~\cite{feng2023promptmagician,brade2023promptify}, enabling efficient candidate exploration and prompt refinement. However, translating these approaches to T23D generation introduces unique challenges. Unlike 2D images, 3D models require physical structural integrity. Moreover, effective evaluation must consider geometry, texture, and coherence across multiple views—a multi-dimensional quality assessment problem beyond current prompt engineering frameworks.}

To bridge this gap, \textbf{\textit{can we develop a visual approach that transforms costly trial-and-error into structured exploration for text-to-3D generation?}} Our formative study (\cref{sec:formative_study}) with professional designers reveals key inefficiencies: limited model options, view inconsistency issues, and lack of effective evaluation and iteration mechanisms.

We present \textit{Sel3DCraft}, an interactive visual prompt engineering system for T23D generation. The system features a dual-branch structure for initial candidate synthesis (\cref{sec:multi_visualization}). 
\sidecomment{R2.Q2\\SR.Q2}\revision{It leverages multi-view images as a 3D visual medium and integrates Multi-modal Large Language Models (MLLMs) for multi-dimensional assessment (\cref{sec:multi-scoring}). Notably, the prompt recommendation mechanism (\cref{sec:score prompt recommendation}) follows the \textit{what-you-see-is-what-you-get} paradigm.}
 Furthermore, we propose a cross-modal visual analytics representation (\cref{sec:interact_sel3dcraft}) that integrates three coordinated components:  
(1) multi-view satellite charts for clustering multi-view images via quality metric projection,  
(2) multi-view hybrid-level scoring heatmaps for model evaluation and defect localization, and  
(3) a prompt-driven treemap wordle for diverse semantic recommendation. 

\sidecomment{R2.Q2\\SR.Q2}\revision{Validation of the semantic scoring approach (\cref{sec:tech_evaluation}) and user study (\cref{sec:user_study}) demonstrate \textit{Sel3DCraft}'s potential and superior creativity support. Compared to baselines, \textit{Sel3DCraft} reduces model creation time by 70.5\% (118.83s vs 402.17s) and prompt iterations by 66.2\%, with significantly higher model quality ratings (4.58 vs 2.46, \cref{fig:baseline_questionnaire}).}

Our contributions are threefold:

\begin{itemize}
\item \sidecomment{R2.Q2\\SR.Q2}We introduce T23D visual prompt engineering as a novel task. \revision{\textbf{\textit{For the first time, we leverage MLLMs to evaluate multi-view images of 3D models in Visual Analytics Science and Technology}}. This integration enables iterative refinement and significantly improves both the process and the outcome of 3D generation.}

\item We propose a cross-modal multi-view visual representation approach for T23D generation. A user study validated the usefulness of these visualizations in boosting user engagement.

\item We present \textit{Sel3DCraft}, an interactive visual system unifying retrieval and generation. Through user evaluations, we demonstrate that the system achieves faster model creation while maintaining comparable quality.
\end{itemize}

\section{Related Work}
\subsection{Visual Prompt Engineering}

Prompt engineering has evolved significantly with advances in large language models \cite{ouyang2022training, brown2020language} and text-to-image generation systems \cite{nichol2021glide, ramesh2022hierarchical}. Early work focused on automated text-to-text prompt optimization through methods like AutoPrompt \cite{shin2020autoprompt} and template generation \cite{gao2020making}. As tasks increased in complexity, interactive systems like PromptIDE \cite{strobelt2022interactive} and PromptChainer \cite{wu2022promptchainer} enabled collaborative human-AI refinement of prompts.

Visual prompt engineering introduces new challenges for text-to-image models. Studies by Liu \cite{liu2022design} and Oppenlaender \cite{oppenlaender2022prompt} established foundational taxonomies for prompt modifiers. Recent advancements such as Li et al. \cite{10.1109/TVCG.2024.3388514} developed a visual analytics system to guide image captioning and uncover biases, aligning with AI-assisted content creation goals. Wang \cite{wang2023democratizing} demonstrated how visual prompt engineering democratizes content creation, while PromptMagician \cite{feng2023promptmagician} refines prompts through multi-level visualizations.

However, current prompt engineering is largely confined to 2D. To overcome this limitation, we propose a novel approach leveraging multi-view images as an intermediate representation. The key innovation is our semantic-based treemap wordle, which introduces T23D image-text matching \cite{li2022blip} to link prompts to multi-view images. This allows users to \textit{intuitively perceive how different parts of prompts contribute to specific 3D regions} and iteratively refine inputs by selecting relevant keywords.

\subsection{Creativity Support Systems}
Creativity support systems (CSS) have been extensively studied in HCI by leveraging generative models and interactive tools to enhance human creativity in design workflows. In text-to-text and text-to-image generation, systems such as TaleBrush~\cite{2022TaleBrush} combined sketches with generative language models to support storytelling. PromptPaint~\cite{Chung2023PromptPaintST}, on the other hand, enabled users to guide image generation through paint-like interactions, highlighting CSS's potential to inspire both visual and textual creativity. In interface design, Scout~\cite{Swearngin2020ScoutRE} supported rapid exploration of layout alternatives through high-level constraints, underscoring CSS's role in enhancing design efficiency and fostering innovation. For visual search, GenQuery\cite{Son2023GenQuerySE} used generative models to support expressive query formulation, helping users retrieve content aligned with their creative intent. 

In 3D design, 3DALL-E\cite{Liu20223DALLEIT} incorporated text-to-image AI into 3D modeling workflows, providing designers with inspiration and reference images while preventing design fixation. In contrast, our system directly generates diverse and high-quality 3D models, \textit{offering designers immediate creative inspiration rather than relying on indirect image references}.

\subsection{Shape Exploration}
The field of shape exploration has evolved significantly, driven by the need to navigate and understand large collections of images and 3D models. The pioneering work of Design Galleries by Marks et al.~\cite{marks1997design} laid the foundation by introducing a novel user interface concept for browsing collections through parametric representation space. Building on this, Averkiou et al.~\cite{averkiou2014shapesynth} extended the concept into 3D space, advancing shape exploration methodologies.
Subsequent research introduced techniques like tree-based visualizations~\cite{bertucci2022dendromap}, augmented scatter plots~\cite{xia2022interactive, zhu2021visualizing, wang2021m2lens, wong_anchorage_2023}, and node-link graphs~\cite{zeng2019emoco,zeng2022gesturelens,liang2022multiviz} for large-scale exploration. These approaches have enriched our ability to visualize and delve into complex datasets, revealing intricate structures and relationships.
In interactive summarization and conceptual design, Juxtaform by Pandey et al.~\cite{pandey2023juxtaform} enables designers to efficiently navigate large collections and derive new shape ideas, addressing key challenges in conceptual design. 

Existing shape exploration techniques often neglect the dynamic exploration of 3D models generated from textual descriptions and the semantic impact of prompts.
The advent of large-scale 3D datasets~\cite{objaverseXL} has enabled new exploration methods, with the Contrastive Language-Image Pre-training (CLIP) model~\cite{liu2023openshape} bridging modalities in these datasets. Our system utilizes OpenShape, a text-image-3D pre-training model, as the retrieval branch. By leveraging a CLIP-based prior, the system \textit{integrates multi-modal data from text, images, and 3D models, efficiently retrieving 3D models from a vast dataset}.

\subsection{Text-to-3D Generation}
Text-to-3D generation combines linguistic descriptions with advanced rendering techniques to create 3D content directly from textual prompts. Early approaches like DreamFusion~\cite{poole2022dreamfusion} leveraged CLIP~\cite{radford2021CLIP} and NeRF~\cite{mildenhall2020nerf} for text-aligned 3D content, while techniques such as Score Distillation Sampling (SDS)~\cite{wang2023score} optimized 3D representations through 2D diffusion priors.
Recent advancements have addressed three key challenges in T23D generation: representation efficiency, generation speed, and multi-view consistency. For representation, diffusion models have been applied to various 3D formats including point clouds~\cite{nichol2022point}, voxel grids~\cite{tang2023volumediffusion}, 3D Gaussians~\cite{zhang2024gaussiancube,he2024gvgen}, and compact latent spaces~\cite{rombach2022high} with methods like Michelangelo~\cite{zhao2024michelangelo} and LION~\cite{vahdat2022lion}. For generation speed, triplane-based methods like LRM~\cite{hong2023lrm} and Instant3D~\cite{li2023instant3d} offer rapid one-step generation, while Trellis~\cite{xiang2024structured} introduced a hierarchical latent model for high-quality one-shot generation. For multi-view consistency, MVDream~\cite{shi2023mvdream}, Wonder3D~\cite{long2023wonder3d}, and One-2-3-45~\cite{liu2023one} have significantly improved coherence across viewpoints, while LGM~\cite{tang2024lgm} established new efficiency benchmarks.

Unlike these methods that focus primarily on generation algorithms, our system emphasizes the human-in-the-loop exploration experience. We adopt TripoSR~\cite{hong2023lrm} for its efficiency and integrate multi-view MLLM-as-a-Judge evaluation~\cite{mllmasjudge,gpt4v3d} to filter low-quality content  with interactive visual prompt engineering.

\section{Formative Study}
\label{sec:formative_study}
To understand how 3D designers use current text-to-3D tools in 3D content creation, we conducted a formative study involving group interviews (\cref{sec:interview_process}). Through the interview results, we captured the users' insights in the creative workflow (\cref{sec:interview_result}) and derived the design requirements for our system (\cref{sec:design_requirements}). 
\subsection{Interview Process}
\label{sec:interview_process}
The \textbf{target users} of this study are 3D designers who utilize T23D tools for creativity support and inspiration. We recruited three 3D design domain experts: a digital media art professor (U1), a 3D modeler (U2), and a 3D product designer (U3). They all have over ten years of experience in 3D model design and are familiar with T23D platforms for digital content production. During the study, participants were encouraged to freely create using existing text-to-3D websites, such as Meshy\footnote{\url{https://www.meshy.ai}} and Luma AI\footnote{\url{https://lumalabs.ai}}, until they were satisfied with their results. Subsequently, we conducted a 40-minute interview with each participant, focusing on their workflows and the challenges they encountered. 

\subsection{Interview Results}
\label{sec:interview_result}
Based on user feedback from the interviews, we identified four key insights, corresponding to the stages of model design: model selection, macro-level and micro-level model evaluation, and model iteration. 
\begin{itemize}
\item \textbf{Insight 1. Limited model selecting options.} 
In the formative study, users noted the number of models generated in one iteration is limited. U3 mentioned, ``\textit{Since I can only get four models at a time, sometimes I am not satisfied with any of them and have to waste an iteration.}'' U2 added, ``\textit{Having so few choices in one iteration makes the experience frustrating. It doesn't allow for much exploration and feels like a lot of work repeating. }''
\item \textbf{Insight 2. Inconsistency issues between model views.}
Unlike 2D images, generated 3D models often appear correct from the front but deform on the sides and appear dark on the back. This is a common issue in generative models known as view inconsistency. In current T23D tools, users need to manually rotate them to observe inconsistencies. U1 stated, ``\textit{Manually rotating each model is a bit annoying and I can't immediately identify flaws for complex models.}'' 

\item \textbf{Insight 3. Requirements for intuitive model evaluations.}
When selecting a satisfactory model, users need to individually evaluate each model. Intuitive evaluation assistance can save time and improve efficiency, even for experienced users. U3 emphasized, ``\textit{Clear, objective metrics for geometry cleanliness or texture fidelity would significantly streamline the decision-making process.}'' 

\item \textbf{Insight 4. Difficulty in crafting appropriate text prompts.}
Users often fail to achieve desired results from text input in a single iteration. However, they also find it challenging to effectively refine prompts in subsequent iterations. U3 noted, ``\textit{I'm not that sure about what level of detail is required in the prompt to guide model generation properly, especially when controlling specific factors.}''  U1 shared, ``\textit{If the system could suggest better ways to phrase my inputs, it would save me a lot of trial and error.}'' 
\end{itemize}
\subsection{Design Requirements}
\label{sec:design_requirements}
Based on the insights from the interview results, we derive the following design requirements: 
\begin{itemize}
\item \textbf{R1. Generate enough images for exploration and selection (Insight 1).}
As stated in Insight 1, most websites typically provide only three to five model candidates per iteration. This severely restricts the users' exploration space, forcing them to undergo multiple rounds of iteration. Therefore, the system should provide a sufficient number and a wide variety of image candidates for users to browse, helping them efficiently find preferred images.

\item \textbf{R2. Support 3D-aware multi-view results (Insight 2).}
Although rotating a single model is not time-consuming, individually rotating each model for observation is a waste of time. To reduce observation time, the system should support a multi-view exploration mode that highlights inappropriate views, enabling users to quickly filter models.

\item \textbf{R3. Provide human-aligned multi-level evaluation (Insight 3).}
When selecting models, subjective aesthetic standards are determined by the 3D designers themselves. However, objective evaluation criteria, such as lighting and color consistency or physical structure, can be automated to save time. We find it essential to provide automatic evaluation aligned with human semantics, helping users efficiently identify and discard low-quality models.

\item \textbf{R4. Recommend iterative refinement of prompts (Insight 4).} 
 When users modify text prompts, numerous manual attempts may not yield the desired results, resulting in a waste of time. If the system provides prompt suggestions after each iteration, guiding users towards discovering phrases that can further boost model output, it would not only help users find satisfactory models but also improve model generation efficiency. 
\end{itemize}

\section{Technical Method}
\sidecomment{R2.Q2\\SR.Q2}\revision{Our approach directly addresses the design requirements (R1-R4) through a tightly integrated pipeline. This pipeline combines multi-modal exploration, perceptual evaluation, and interactive refinement. \textit{Sel3DCraft} transforms the conventional T23D workflow from an unguided trial-and-error process into a structured visual exploration. This transformation is guided by human perceptual metrics.}

\begin{figure*}[ht]
 \centering
 \sidecomment{R1.Q3\\SR.Q3}
 \revisionbox{\includegraphics[width=\linewidth]{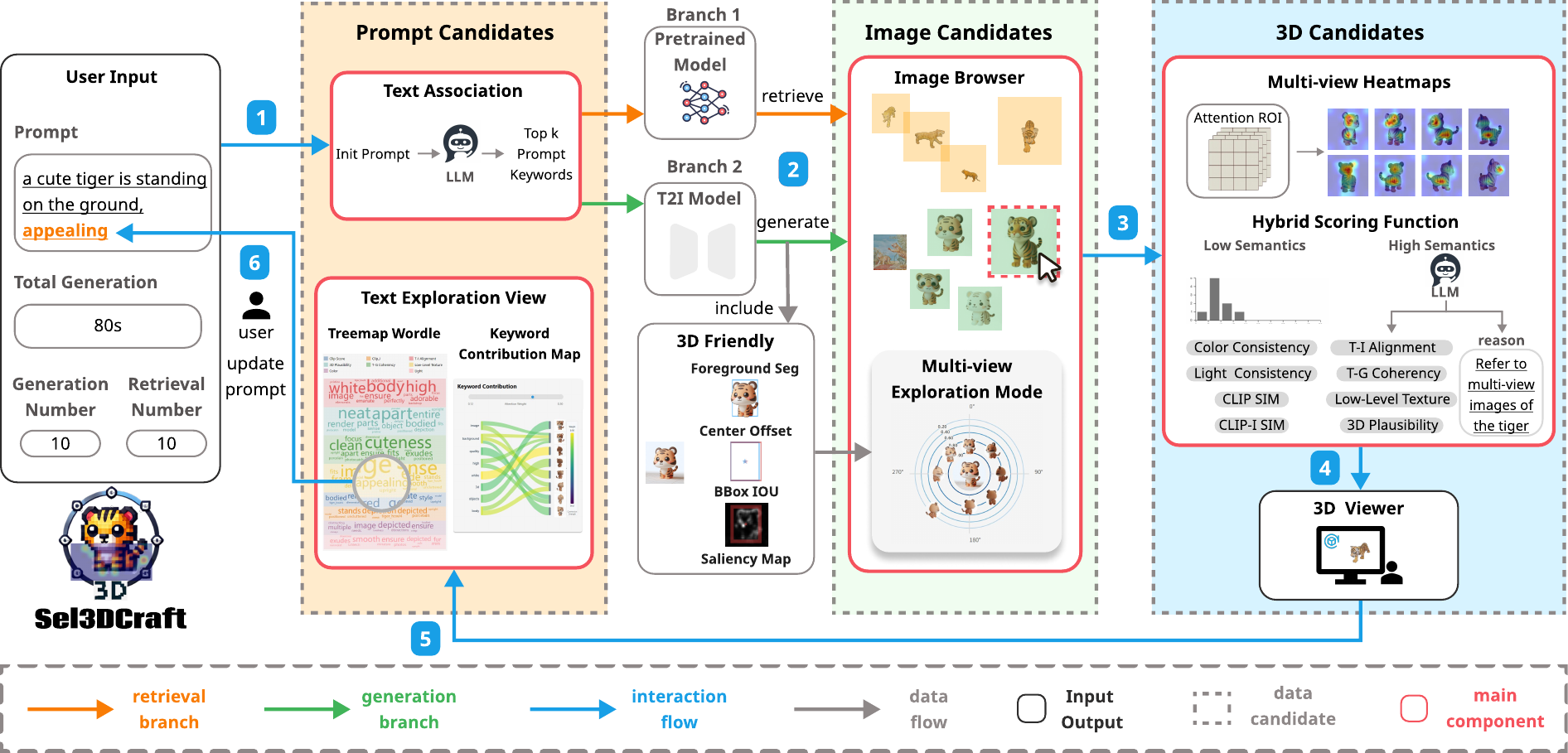}}
 \caption{
 The \textit{Sel3DCraft} pipeline begins by expanding user text input into multiple candidates via LLM. A dual-branch pipeline then generates/retrieves corresponding multi-view images, displayed in an interactive satellite view for 3D perception. Each candidate is automatically evaluated using eight-level semantic scores with heatmap visualization. For refinement, a score-guided recommendation mechanism suggests optimal keywords, enabling iterative optimization until user satisfaction is achieved.
}
 \label{fig:framework}
 \vspace{-14pt}
\end{figure*}

\sidecomment{R2.Q2\\SR.Q2\\R3.Q6}\revision{The framework (\cref{fig:framework}) implements a three-stage pipeline. The exploration stage expands the candidate space through parallel retrieval and generation pathways (\textbf{R1}). A large language model generates semantically diverse prompt candidates $T_{P}$. Simultaneously, a text-to-3D pretrained model retrieves candidate shapes $S_{3D}$ and the T2I model generates image sets $I_{T2I}$. This creates a rich design space for exploration.}
\sidecomment{R2.Q2\\SR.Q2}\revision{The evaluation stage addresses multi-view consistency challenges (\textbf{R2}) through comprehensive perspective rendering and automated assessment. Both low-level $F_l(\cdot)$ and high-level $F_h(\cdot)$ scoring functions serve as subcomponents of our multi-view scoring function $F_{mv}$. These functions provide human-aligned quality metrics (\textbf{R3}) that identify structural and semantic inconsistencies without manual inspection.}
\revision{The refinement stage translates evaluation metrics into visual guidance for prompt engineering (\textbf{R4}). We project quality scores onto visual attributes to create perceptually intuitive representations. These representations include a prompt treemap wordle and keyword contribution map. This enables users to make informed adjustments based on quality feedback rather than blind iteration.}

\subsection{Multi-modal Unified Visual Representation}
\label{sec:multi_visualization}
\sidecomment{R2.Q2\\SR.Q2}\revision{At the core of our T23D visual prompt engineering framework is a multi-modal unified retrieval mechanism. This mechanism bridges the semantic gap between users' textual descriptions and the desired 3D output. Our approach integrates text, 2D images, and 3D shape representations into a coherent exploration space. This allows users to understand the relationships between separate spaces (\textbf{R1}).}

\textbf{Pre-trained Model-based Retrieval.}
\sidecomment{R2.Q2\\SR.Q2}\revision{Since there is no specific dataset like diffusionDB~\cite{wang2022diffusiondb} for T2I, efficient retrieval from large 3D model repositories is crucial for T23D Visual Prompt Engineering. Mainstream T23D models often use the Objaverse dataset, containing over 800,000 annotated 3D models. To facilitate retrieval, we use OpenShape~\cite{liu2023openshape}, a large-scale Pretrained Model built on Objaverse. OpenShape integrates text, images, and 3D shapes into joint embeddings. Using OpenShape, our pipeline retrieves prompt candidates $T_{PC}$, image candidates, and 3D shape candidates $S_{3D}$ for each set of prompt inputs.}

\revision{\textbf{Multi-modal Synthesis.} The second branch first gets prompt candidates $T_{T2I}^{}$ by augmenting the retrieval prompt $T_{PC}$ using an association module. It then generates image candidates $I_{T2I}$ applying a text-to-image (T2I) model~\cite{podell2023sdxl}. The generated images are assessed by a 3D-friendly scoring function $F_{3Df}$ to select the most suitable image candidates for 3D modeling. TripoSR\cite{hong2023lrm} generates 3D models from the selected images and converts them to multi-view renderings $I_{mview}$. The renderings are evaluated by a multi-view scoring function $F_{mv}$ to compute keyword scores.}

\textbf{Integration and User Interaction.}
The fusion step combines the outputs from two distinct branches within a joint embedding space:
\begin{equation}
    F_{fs}(T, I, S) = \mathcal{Z}(T_{PC}, T_{T2I}^{*}, S_{3D}^{*}),
\end{equation}
where $F_{fs}$ denotes the fusion function, $T$ the input text, $I$ the input image, $S$ the input 3D model, $\mathcal{Z}$ the joint embedding space for image browser, $T_{PC}$ the text from retrieval, $T_{T2I}^{}$ the augmented text for T2I, and $S_{3D}^{}$ the optimized 3D model.

\textbf{\textit{The key technical challenge lies in establishing effective cross-modal alignment between textual descriptions, 2D images, and 3D models while preserving modality-specific information.}} Our joint embedding construction process addresses this through a two-stage approach: first aligning text-image pairs through contrastive learning, then extending to 3D space via multi-view rendering representation. 
Further implementation details are provided in Appendix A.3.

\textbf{3D-Friendly Score.}
Since not all 2D images are suitable for generating 3D models, we introduce a metric to evaluate '3D-friendliness' in multi-view image mode. \sidecomment{R1.Q1\\SR.Q1}\revision{Through empirical analysis of 200 images and their corresponding 3D models, we found that images containing multiple objects within complex backgrounds consistently produced incomplete or distorted 3D models, with a 78\% failure rate observed across our test set. We discovered a strong correlation between foreground segmentation quality and 3D model completeness (Pearson's r=0.82), with centered, single objects against simple backgrounds achieving 3.4 times higher quality scores than complex scenes.}

The metric integrates three key components: Center Offset (CO), Bounding Box Intersection Over Union ($IoU_{BB}$), and binary map Intersection Over Union (IoU), leveraging saliency maps \cite{hou2007saliency} and SAM \cite{kirillov2023segany} for foreground segmentation. The formula for our composite metric $f_{3Df}$ is given by:
\begin{equation}
f_{3Df}(I) = w_O \cdot (1 - \frac{O}{O_{max}}) + w_{IoU} \cdot IoU + w_{BB} \cdot IoU_{BB}.
\end{equation}

\sidecomment{R1.Q1\\SR.Q1}\revision{We tested various weight combinations on a validation set of 50 images and evaluated the correlation between our score and human expert ratings of 3D model quality. Through this process, we found that the final weights  $w_O=0.3$, $w_{IoU}=0.4$, and $w_{BB}=0.3$  achieved the highest correlation coefficient (r = 0.76) with expert assessments.} This weighting reflects our finding that segmentation quality (IoU) most strongly impacts generation outcomes, while center positioning and bounding box coverage are also significant but less influential. 
\subsection{Multi-view Hybrid-level Scoring Function}
\label{sec:multi-scoring}
\sidecomment{R2.Q2\\SR.Q2}\revision{While our dual-branch architecture expands the exploration space, this advantage becomes meaningful only when coupled with efficient filtering mechanisms. Addressing \textbf{R2} and \textbf{R3}, we developed a hybrid scoring framework. This framework combines low-level computational metrics with high-level perceptual judgments through MLLM-based semantic evaluations. Through systematic literature analysis and iterative refinement with 3D artists, we identified eight quality dimensions. These dimensions are structured as complementary semantic hierarchies. Low-level semantics quantify measurable properties including light-color consistency and cross-view coherence via CLIP-based metrics. High-level semantics leverage MLLM perception to evaluate abstract qualities. These include 3D plausibility, text-image alignment, texture-geometry coherency, and low-level texture. Table \ref{tab:scoring_metrics} summarizes our eight semantic scoring dimensions.}

\subsubsection{Low-level Scoring Function}
\label{sec:lowlevel}
To address inconsistency issues, we propose a low-level scoring function aimed at assessing image suitability for reconstruction.

\begin{table}[h]
\centering
\sidecomment{R2.Q1\\SR.Q1}
\revision{
\caption{Eight-Dimensional Semantic Scoring Framework for Multi-view 3D Model Evaluation}
\label{tab:scoring_metrics}
\begin{tabular}{@{}p{2.5cm}p{1.0cm}p{4.2cm}@{}}
\toprule
\textbf{Metric} & \textbf{Inputs} & \textbf{Role \& Computational Method} \\
\midrule
\multicolumn{3}{l}{\textit{Low-Level Metrics}} \\
\midrule
Color Consistency & $I_{mview}$ & Consistency assessment via Bhattacharyya coefficient on Lab histograms~\cite{kailath1967divergence} \\
Light Consistency & $I_{mview}$ & Lighting evaluation via L-channel histogram comparison~\cite{mildenhall2020nerf} \\
CLIP Score & $(T, I)$ & Semantic alignment via text-image cosine similarity~\cite{hessel2021clipscore} \\
CLIP-I Score & $(I_{ref}, I_j)$ & Cross-view coherence via feature alignment using CLIP~\cite{radford2021learning} \\
\midrule
\multicolumn{3}{l}{\textit{High-Level Metrics}} \\
\midrule
Text-Image Alignment & $(T, I_{mview})$ & Semantic correspondence via MLLM-based evaluation~\cite{gpt4v3d} \\
3D Plausibility & $I_{mview}$ & Spatial reasoning via MLLM-based assessment~\cite{mllmasjudge} \\
Texture–Geometry Coherency & $I_{mview}$ & Surface analysis via MLLM-based evaluation~\cite{gpt4v3d} \\
Low-Level Texture & $I_{mview}$ & Detail assessment via MLLM-based evaluation~\cite{mllmasjudge} \\
\bottomrule
\end{tabular}
}
\end{table}

\textbf{Color and Light Consistency.} 
Multi-view consistency is a core challenge in 3D generation tasks~\cite{mildenhall2020nerf}, and the most intuitive visual manifestation of multi-view consistency is the consistency of color and light.
Ensuring color and light consistency $f_{color}, f_{light} \in F_l$ across multi-view images $I_{mview}$ generated from 3D candidate models $S_{3D}$ is crucial. 
\sidecomment{R1.Q1\\SR.Q1}\revision{We analyze Lab color histograms and L-channel histograms across multi-view images to evaluate consistency. The Lab color space aligns closely with human visual perception, making it ideal for assessing color consistency. We use the Bhattacharyya coefficient to compare individual Lab histograms against the average, with values closer to 1 indicating higher consistency.}
\sidecomment{R2.Q1\\SR.Q1}\revision{Specifically, for color consistency we compute $f_{color} = \frac{1}{N} \sum_{i=1}^{N} BC(H_{Lab}^{(i)}, \bar{H}_{Lab})$, where $N$ is the number of views, $H_{Lab}^{(i)}$ is the Lab histogram of view $i$, and $BC$ denotes the Bhattacharyya coefficient. Light consistency follows the same approach using L-channel histograms.}

\textbf{CLIP Score vs CLIP-I.} \sidecomment{R2.Q1\\SR.Q1}Beyond visual attributes, multi-view consistency also encompasses semantic and feature-level coherence. To satisfy designers’ requirements for 3D models to accurately reflect user intent and maintain consistency across different views, we have selected two metrics: CLIP Score~\cite{hessel2021clipscore} and CLIP-I~\cite{radford2021learning}. \revision{CLIP Score measures text-image semantic alignment using cosine similarity between prompt embeddings and image embeddings in CLIP space, formulated as $f_{clip}(T, I) = \cos(E_{text}(T), E_{image}(I))$. In contrast, CLIP-I assesses internal visual consistency by comparing the front view (0° rotation) against other viewpoints of the same 3D model, calculated as $f_{clip-I}(I_{ref}, I_j) = \cos(E_{image}(I_{ref}), E_{image}(I_j))$ where $I_{ref}$ is the reference front view.} 


\subsubsection{High-level Scoring Function} 
While low-level scores measure visual consistency through direct attribute computation, high-level semantic evaluation employs MLLMs to assess results from human perspectives. Compared to traditional evaluation methods, MLLM-as-a-Judge ~\cite{gpt4v3d,mllmasjudge} provides explanatory evaluations with comprehensive performance feedback, serving as a scalable and reproducible alternative to costly manual assessment.

According to discussions with designers, we involve four human-aligned scores based on R3: 
\textbf{Text-Image Alignment}, \textbf{3D Plausibility}, \textbf{Texture-Geometry Coherency}, and \textbf{Low-Level Texture}. 

\sidecomment{R2.Q1\\SR.Q1}\revision{These metrics are precisely defined as follows: \textbf{Text-Image Alignment} evaluates semantic correspondence between input prompts and generated models using MLLM reasoning beyond simple CLIP similarity. \textbf{3D Plausibility} assesses spatial coherence and physical realism across viewpoints, determining whether the model maintains structural integrity in 3D space. \textbf{Texture-Geometry Coherency} examines how surface details align with object contours and shape, ensuring realistic texture mapping. \textbf{Low-Level Texture} focuses on fine detail preservation including edges, patterns, and color variations that contribute to visual fidelity.}

\sidecomment{R1.Q1\\SR.Q1}\revision{Our MLLM prompt design includes several key components: clear task definitions specifying evaluation criteria and scoring scales; contrastive examples presenting paired high- and low-quality outputs for calibration; and a step-by-step evaluation framework guiding the model through structured assessment. 

We use paired images for contrastive scoring, where the MLLM compares consecutive views of the same 3D model to generate both semantic scores and explanatory reasoning. This approach provides transparent, step-by-step evaluations that enhance user understanding of model quality assessments. The complete prompt template can be found in Appendix C.2.}


\subsection{Score-guided Prompt Recommendation}
\label{sec:score prompt recommendation}



The semantic gap between text and 3D outputs hinders text-to-3D creation, as users struggle to convert quality feedback into actionable prompts. We address this through a visual recommendation system that translates abstract quality assessments into concrete semantic suggestions (\textbf{R4}).

\sidecomment{R1.Q6}\revision{Traditional prompt engineering lacks structured guidance. Our treemap wordle bridges this gap by mapping quality metrics to keywords through spatial layout, size, and color.}

\textbf{Prompt Conceptual Augmentation.} To enhance the text-to-3D synthesis, we introduce a streamlined prompting mechanism designed to refine and extend the semantic scope of user inputs using LLMs. This is formally expressed by:
\begin{equation}
T_P = f_{aug}(t_0),
\end{equation}
where $f_{aug}(\cdot)$ denotes the mapping function that combines the original prompt  $t_0$ with a predefined corpus of 3D-specific prompts to generate a set of enriched prompt candidates $T_P = \{t_1, \ldots, t_N\}$.
 
\textbf{Treemap Wordle Prompt Recommendation.} \sidecomment{R2.Q2\\SR.Q2}\revision{An essential aspect of our visual prompt recommendation system is to capture and adapt to users' preferences for different criteria of 3D model generation. The multi-view scoring function $F_{mv}$ evaluates each 3D model $s \in S_{3D}$ from eight perspectives. This yields a score vector $F_{mv}(s) \in \mathbb{R}^8$. To provide recommendations that align with users' interests, we propose a treemap wordle visualization for interactive prompt exploration.}
\revision{The key idea is to project the high-level score vectors into a visually interpretable low-dimensional space. In this space, similar prompts are grouped together and important keywords are highlighted.}

\textbf{Prompt Clustering and Keyword Extraction.} \sidecomment{R1.Q2}\revision{We apply HDBSCAN clustering on 2D embeddings to group similar prompts. For each cluster, we compute word frequencies and select the most frequent words as representative keywords. The score of each keyword is calculated as the average of all scores from models containing that keyword, enabling quality-based keyword recommendations.}

\sidecomment{R1.Q6}\revision{\textbf{Visual Encoding.} We visualize clusters using a treemap layout where rectangular areas reflect cluster sizes. Keywords appear as word clouds with font sizes indicating frequencies and colors representing eight-level semantics (see \cref{fig:data_flow}).}

\revision{\textbf{Keyword Contribution Map.} We compute CLIP cross-attention maps to connect keywords with multi-view images using IoU between attention maps and foreground regions.} 

\sidecomment{R2.Q2\\SR.Q2}\revision{The treemap wordle visualization highlights dominant themes in generated 3D models by prevalence and quality. By clicking on a keyword $w_{ki}$, the user can filter 3D models to include only those models $s$ whose prompts $p_s$ contain $w_{ki}$. This allows targeted exploration of specific preferences. The user can also merge multiple keywords to compose new prompts on the fly. Throughout the prompt update process, the visual prompt recommendation system updates the treemap wordle dynamically. This reflects the user's current focus.}

\section{Sel3DCraft System}
This section introduces \textit{Sel3DCraft}, an interactive visual analytics system that supports dual-branch 3D model generation from textual descriptions, and provides interactive features such as model evaluation and prompt recommendations. In \cref{sec:user_interface}, we present the system's interface design. We then explain how users interact with the system (\cref{sec:interact_sel3dcraft}) and demonstrate the interaction through scenario design (\cref{sec:scenario_design}). 

\subsection{User Interface}
\label{sec:user_interface}
Our system's design philosophy focuses on maximizing users' ability to perceive and understand relevant information with minimal interaction. 
As shown in \cref{fig:teaser}, the system's interface comprises five modules: (A) Model Input View, enabling users to input initial prompts and perform prompt associations (\textbf{R1}); (B) Image Browser View, which displays images retrieved and generated from prompts as satellite charts (\textbf{R2}), and presents multi-view heatmaps along with eight-level semantic scores (\textbf{R3}); (C) 3D Viewer View, showcasing the final generated model for observation; (D) Text Exploration View, clustering prompts by semantics into a treemap wordle to recommend keywords for iterative modifications (\textbf{R4}); and (E) Keyword Contribution View, displaying the semantic contribution of keywords to multi-view images (\textbf{R4}). All modules are closely interconnected through multi-modal conversions of text, images, and 3D models. The following sections detail the innovative visualization designs within the interface.

\subsection{Interacting with Sel3DCraft}
\label{sec:interact_sel3dcraft}
\begin{figure}
    \centering
    \includegraphics[width=1.0\linewidth]{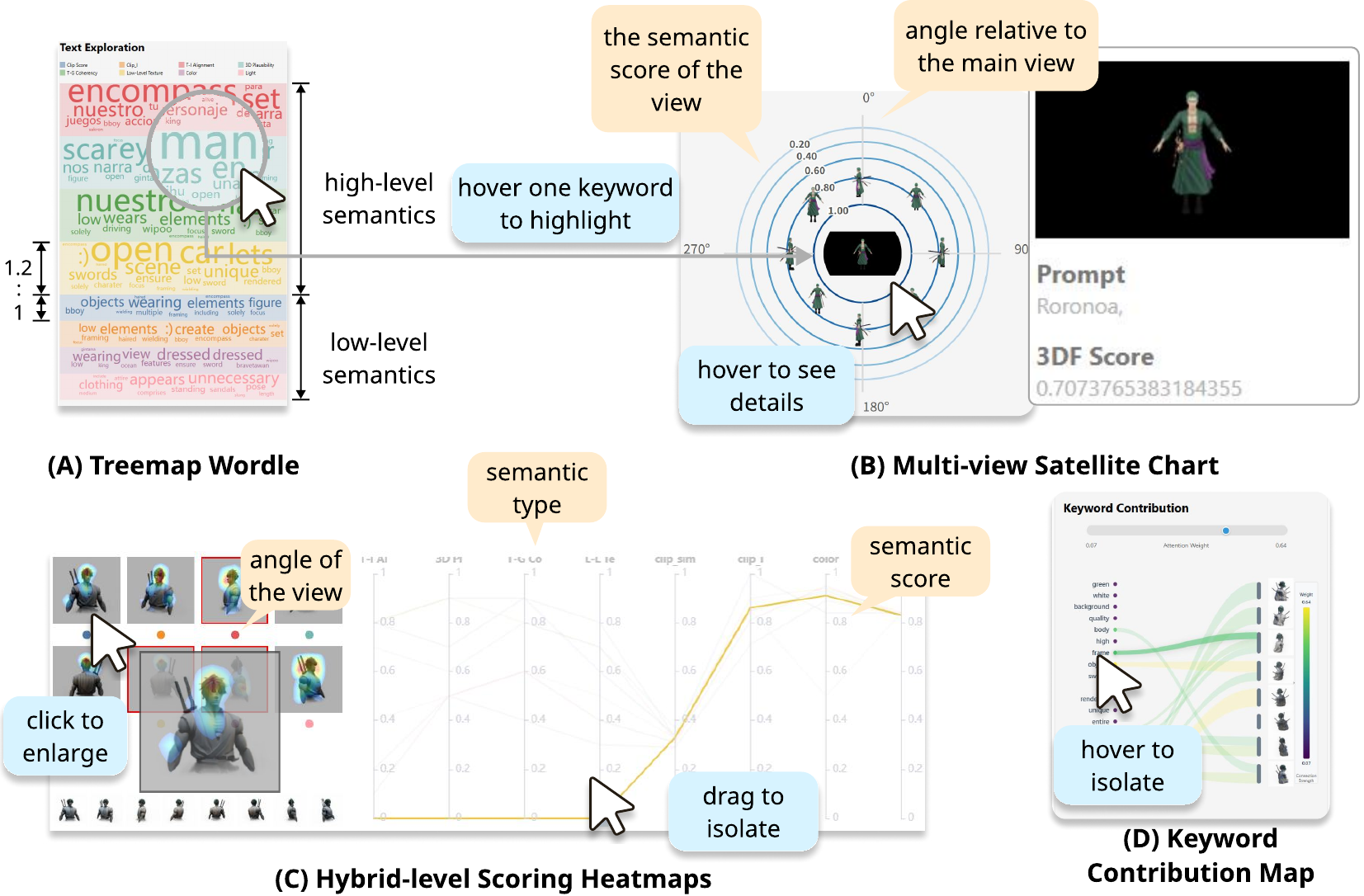}
    \caption{Interactive design of visualization charts and mapping relationships between data. (A) illustrates the visual design of the treemap wordle. (B) depicts the design of the multi-view satellite chart. (C) presents the system's multi-view hybrid-level scoring heatmaps. (D) shows the keyword contribution map representing the semantic contribution.
    }
    \label{fig:data_flow}
            \vspace{-14pt}
\end{figure}

As shown in \cref{fig:data_flow}, the interactive modules of this system consist of four main components: multi-view satellite charts, hybrid-level scoring heatmaps, a treemap wordle, and a keyword contribution map.  Next, we will provide a detailed introduction to each of these components (For a more intuitive view of the interactions, please refer to our supplementary video). 

\subsubsection{Multi-view Images Surrounding Satellite Chart}

The satellite chart's multi-view display enables intuitive browsing of prompt-generated images, allowing users to quickly identify and discard those with perspective defects without needing to rotate and inspect each one in the 3D viewer. As shown in \cref{fig:data_flow} (B), the central image shows the frontal view, while eight surrounding images (at 45° intervals) represent rotation angles (0-360°) of multi-view images relative to the main view. \sidecomment{R1.Q7\\SR.Q3}\revision{The distance of each view from the center reflects its semantic score for 3D plausibility—lower-scoring views are positioned farther outward, with circle colors radiating from deep blue to light blue.} Therefore, a tighter cluster indicates higher overall quality. Users can hover to view full prompts and individual scores (\textbf{R2}).

\subsubsection{Multi-view Hybrid-level Scoring Heatmaps}
The Image Browser View's bottom panel presents hybrid-level scoring through two visual components  (\cref{fig:data_flow} (C)). First, heatmaps highlight defect areas in red, with heavily flawed images outlined in red frames—users can click for detailed inspection. Second, a multi-line chart tracks eight semantic dimensions (x-axis) versus their scores (y-axis), with each colored line representing a view. By long-pressing the left mouse button on the vertical axis, users can create a selection to isolate and view semantic polylines within the selected area, providing a more intuitive and focused view (\textbf{R3}).

\subsubsection{Treemap Wordle Text Exploration}
The treemap wordle (\cref{fig:data_flow} (A)) clusters text from all 3D candidates to recommend supplementary keywords. Its eight horizontally divided color-coded sections represent different semantics, with high-level semantics (deemed more important by designers) occupying 20\% more area than low-level ones. Higher-frequency keywords appear larger within each section for quick selection. When a user hovers over a keyword, related images in the Image Browser are highlighted. Adding keywords to the input triggers a new iteration, regenerating both image and 3D candidates. 
\sidecomment{R1.Q7\\SR.Q3}\revision{The keyword contribution map (\cref{fig:data_flow} (D)) employs a Sankey diagram to visualize each keyword’s impact to the multi-view images. Contribution values are encoded using a color gradient from yellow to purple, representing high to low contributions, respectively. A threshold slider allows users to filter out lower-contributing links, retaining only the most significant connections (\textbf{R4}).}

\subsection{System Scenario Design}
\label{sec:scenario_design}
The system supports three main usage scenarios. In Scenario 1, users create the character Cattiva from "PalWorld,"\footnote{\url{https://palworld.gaming.tools/en}} stopping when the model sufficiently resembles the target. This demonstrates the system's accuracy and quality in matching specific styles. \sidecomment{R3.Q1\\SR.Q4}\revision{In Scenario 2, users design a logo for \textit{Sel3DCraft} with creative freedom in visual style, which showcases the system's ability to foster creativity while contrasting with Scenario 1's replication task.} 
\sidecomment{R3.Q2\\R3.Q5\\SR.Q4}\revision{In Scenario 3, users generate teacup models with watertight geometry and sufficient wall thickness to ensure 3D printability, highlighting the system's practical applicability and fabrication potential. }

\begin{figure}
    \centering
     \sidecomment{R1.Q5\\SR.Q3}
      \revisionbox{\includegraphics[width=1.0\linewidth]{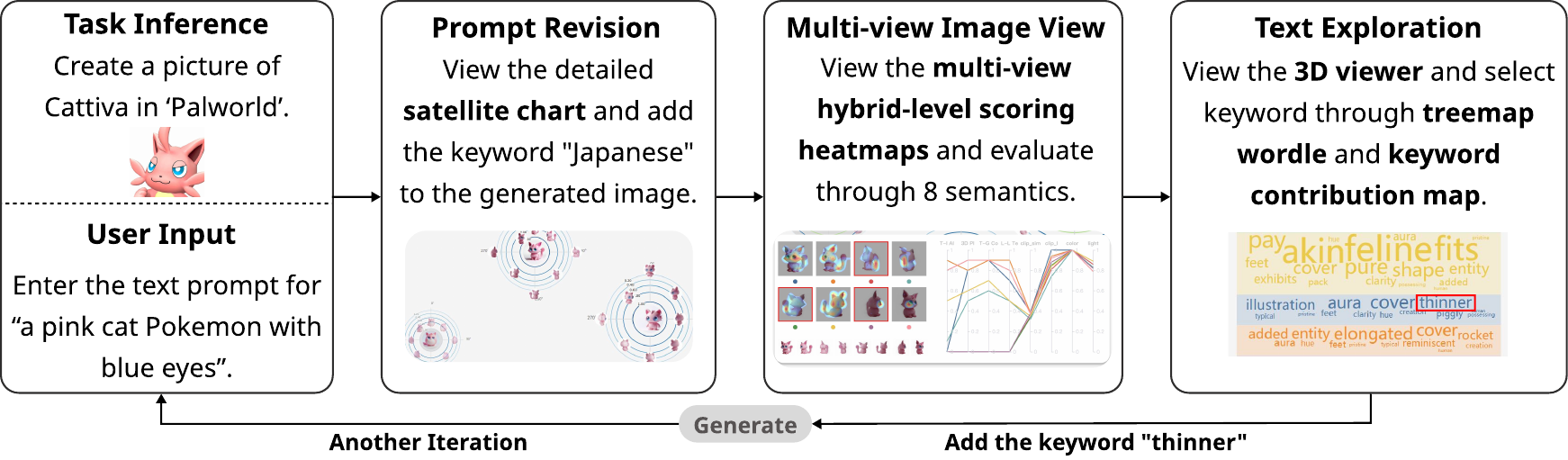}}
    \caption{The first scenario demonstrates the system's ability to generate targeted models. 
    }
    \label{fig:pal_design}
\end{figure}

Take Scenario 1 as an example. The interaction flow of the system is illustrated in \cref{fig:pal_design}. The user started with the prompt, ``a pink cat Pokemon with blue eyes,'' to describe Cattiva. Then she selected a preferred multi-view satellite image to view the expanded prompt and added ``Japanese'' as a new keyword. After the system produced more accurate images, the user used the hybrid-level scoring function to select the most satisfactory model from a large pool of candidates. However, by preliminary reviewing in the 3D viewer, the user decided to explore the treemap wordle for further refinement. The user identified ``thinner'' as a significant contributor aligned with her improvement goals while analyzing the keyword contribution map. She then incorporated this term into the prompt and proceeded to another iteration. The final model matched Cattiva's original appearance, demonstrating the system's effectiveness in accurate character modeling.

\revision{\section{Validation of the Semantic Scoring Approach
\label{sec:tech_evaluation}
}}\sidecomment{R1.Q8}

\subsection{Human Evaluation on High-level Semantics}
\begin{figure}
    \centering
    \includegraphics[width=0.8\linewidth]{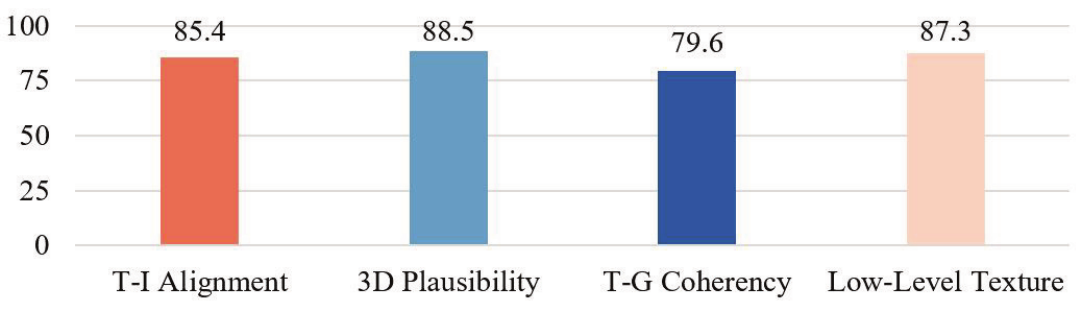}
    \caption{Human evaluation results of high-level semantics on generation.}
    \label{fig:highlevel}
\end{figure}

To further validate the utility of high-order semantics, we measured their alignment with human preferences, as human judgment is often considered the gold standard for quality and reliability. For each order of semantics, evaluators were presented with two sets of multi-view images, one with a higher semantic score based on our system's evaluation mechanism. Evaluators were then asked to select the set with better semantic representation. A response was considered correct if the evaluator selected the set with the higher semantic score. Additionally, we excluded image pairs with semantic score differences of less than 0.3, as humans struggle to distinguish such fine-grained differences.
We invited twenty university students as evaluators to investigate whether non-experts perceive the system's model evaluation as sufficiently accurate. Five-student groups performed single-level semantic evaluation tasks, with each participant distinguishing thirty multi-view image pairs. The results (see \cref{fig:highlevel}) show that evaluators correctly identified the model with higher semantic scores with an accuracy rate of over 79\%. However, consistency in texture-geometry coherency was lower, likely due to the limited familiarity of amateur users with industry standards for 3D models, making such judgments more challenging.
\subsection{Expert Annotation on Semantic Consistency}
We also invited the same three experts (U1-U3) who participated in our formative study to annotate fifty 3D models based on high-level semantics. Subsequently, we employed the Bland-Altman analysis to evaluate the consistency between the average scores from expert annotations and the semantic scores assessed by the MLLMs. The results show that most data points fall within the consistency limits, and the mean difference is close to zero, indicating minimal deviation between the expert annotations and the system-generated scores, thus demonstrating strong consistency (see Appendix D). 

To validate the generalizability of our evaluation system across different AI models, we employed Gemini 2.0 Flash and Qwen2.5VL for automated quality assessment of 3D models, in addition to GPT-4o, which we used for our system's scoring functions. These assessments were compared against expert manual annotations. While minor discrepancies exist in individual samples, all AI evaluation models demonstrate strong alignment with expert ratings in overall trends (see Appendix D). This finding indicates that our cross-model evaluation framework exhibits considerable robustness, where different AI assessors can effectively capture quality metrics analogous to human experts' criteria.



\section{User Study}
\label{sec:user_study}
\sidecomment{R1.Q8\\SR.Q3}\textit{Sel3DCraft}, embodying a human-in-the-loop methodology, necessitates a user study to ascertain its efficacy and usability. \revision{We adopted a multi-stage evaluation strategy. The first phase focused on evaluating (1) the usefulness of the three proposed visualization components, (2) the overall effectiveness of the system, and (3) its creative advantages over two baseline systems. Based on user feedback, we then made minor refinements to the visual styling of the multi-view satellite chart and the keyword contribution map—without altering their core functionality—and carried out a focused follow-up evaluation to assess the impact of these design adjustments.}
\subsection{Participants}
We recruited twelve participants (eight females, four males, average age = 23.7) from various fields, including computer science, digital media, and industrial design. All participants had over three years of professional experience in 3D design and frequently used text-to-3D tools for artistic exploration. On average, they rated their familiarity with text-to-3D tools as 4.75 out of 5. However, their understanding of prompt engineering was relatively limited, with an average rating of 3.76 out of 5. 

\subsection{Baseline Systems}
In the formulation of our study, two baseline systems were designed to serve as comparative benchmarks, enabling a nuanced evaluation of our system's performance and user experience. 

\subsubsection{Vanilla T23D Tool} Baseline A utilizes the Meshy website to generate text-to-3D models. Users can input text, and four preliminary models are generated with each iteration. Users then select the most suitable model for further refinement. However, Baseline A lacks both prompt recommendation and model scoring functionalities. We selected Baseline A as the control group representing the traditional T23D workflow. 

\subsubsection{T2I Visual Prompt Engineering} 
Baseline B employs PromptMagician~\cite{feng2023promptmagician} to retrieve relevant images based on user input and provides real-time keyword recommendations. Users can then convert the selected images into 3D models using the TripoSR\cite{hong2023lrm} web interface provided in our study. Although Baseline B employs prompt engineering similar to our system, it is limited to T2I generation and lacks a corresponding scoring mechanism to validate the feasibility of image-to-model generation. 
\subsection{Procedure and Tasks}

The evaluation process consists of three main phases: introduction, comparative experiment, and survey feedback collection.

 \textbf{Introduction (10 min).}
At the beginning of the evaluation, we introduced the study background and evaluation plan to the users. Next, we described the interface features of the systems and demonstrated their functionalities through clear examples \cite{yang2021explaining}. 

\textbf{Comparative Experiment (30 min).}
The comparative study evaluated three usage scenarios: targeted generation of predefined models, \sidecomment{R3.Q1}\revision{theme-based creative design} based on personal preferences, and preparation of 3D printing-ready models. Participants selected one scenario and used all three systems within ten minutes per system. \sidecomment{R3.Q4\\SR.Q4}\revision{The order of system usage (Baseline A, Baseline B, and \textit{Sel3DCraft}) was randomized across participants to mitigate order effects.}

\begin{figure}[h]
  \centering
  \includegraphics[width=1.0\linewidth]{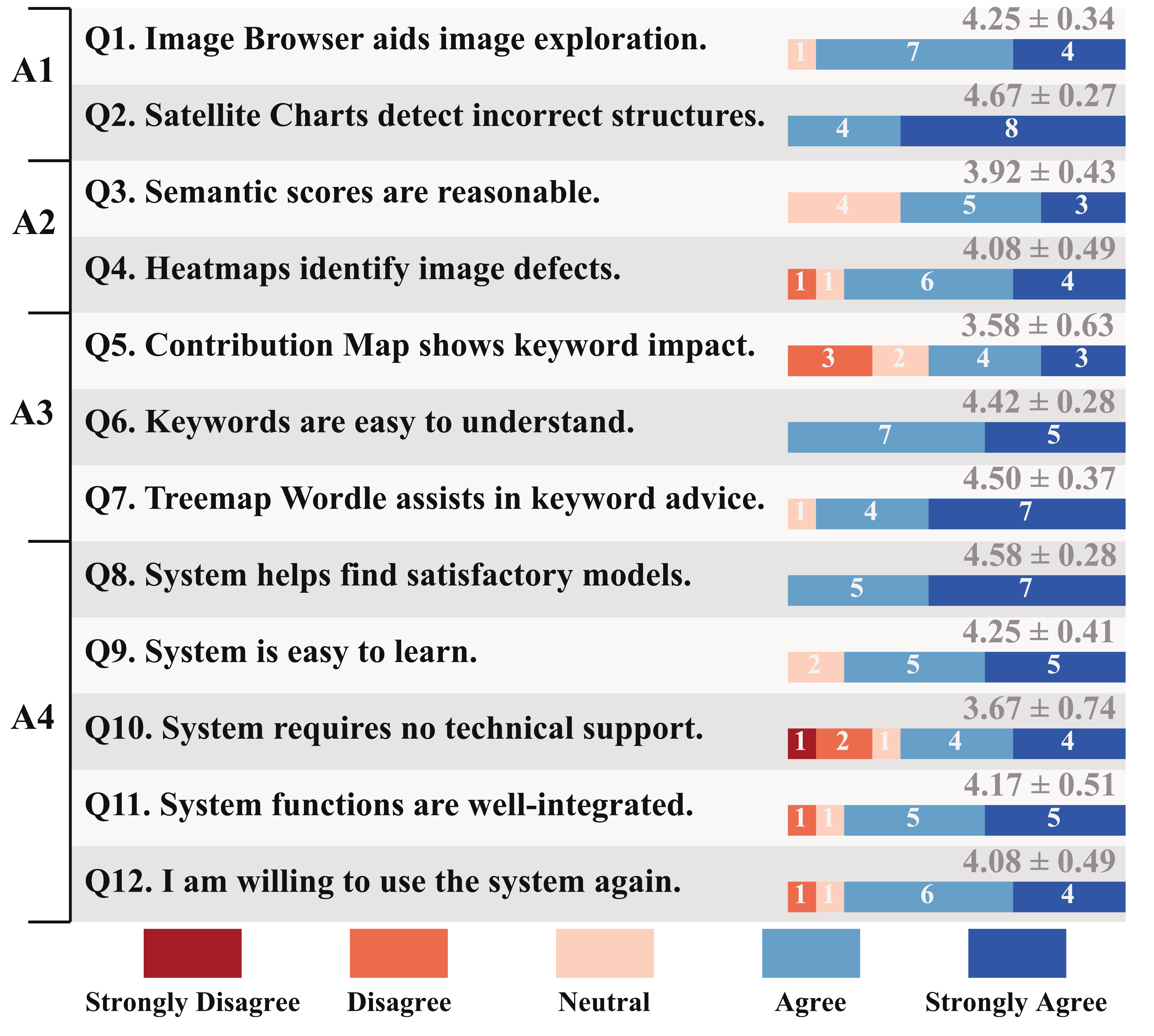}
  \caption{Questionnaire results of the usefulness and effectiveness of our system. \protect\sidecomment{R3.Q3}\protect\revision{A1, A2, and A3 denote usefulness evaluations for the Satellite Chart, Multi-view Hybrid-level Scoring Function, and Treemap Wordle Text Exploration visualizations respectively, while A4 represents the overall system effectiveness assessment.} The average scores and 0.95 confidence intervals are listed above the user option counts on the right side.}
  \label{fig:our_questionnaire}
\end{figure}

 \textbf{Feedback Survey (20 min).}
To conclude the study, we administered two 5-point Likert-scale questionnaires. The first assessed the utility of our system across its three core modules and its overall effectiveness (see \cref{fig:our_questionnaire}). The second evaluated creativity support compared to Baselines A and B (see \cref{fig:baseline_questionnaire}). Follow-up interviews provided qualitative insights for deeper analysis.

\subsection{Results Analysis}
After the comparative experiment, two questionnaires were collected from participants. Based on participant feedback, we analyzed the system's usefulness, effectiveness, and differences compared to the two baseline systems. \sidecomment{R3.Q4\\SR.Q4}\revision{We ultimately divided our results analysis into two parts: qualitative feedback and quantitative comparison to baselines.}
\revision{\subsubsection{Qualitative Feedback}}
\textbf{Usefulness of Satellite Chart}
Most users agreed that the Image Browser provided a \textbf{diverse} range of images (Q1). P1 commented, ``\textit{The view contains a wide variety of image types, allowing for free exploration based on user input.}'' Each single-view image is surrounded by multi-view images in a satellite-like arrangement, helping users identify \textbf{incorrect physical structures} (Q2). P8 noted, ``\textit{When creating a cattiva, I found a misaligned tail in one of the views, so I immediately discarded the image.}''

\sidecomment{R1.Q6}\revision{\textbf{Usefulness of the Multi-view Hybrid-level Scoring Function.} 
The semantic scores were considered \textbf{reasonable} and \textbf{effective} (Q3). P6 easily identified low text-image alignment through the eight-level line chart. Participants agreed that heatmaps effectively highlighted \textbf{defects} (Q4) by marking anomalous areas in red (P12).}

\revision{\textbf{Usefulness of the Treemap Wordle Text Exploration.}
The treemap wordle was widely praised. Users found keyword contributions clearly visualized (Q5) and keywords \textbf{simple} and \textbf{easy} to understand (Q6). P2 noted keywords were primarily clear adjectives and nouns. The Text Exploration View helped users find \textbf{suitable supplementary keywords} (Q7). P4 noted it effectively bridged knowledge gaps during generation.}

\textbf{Effectiveness of the system}
Through iteration, all participants found a \textbf{satisfactory} final model (Q8). Additionally, they found the system \textbf{easy to use} (Q9) and did not require \textbf{technical support} (Q10). The vast majority of participants agreed that the system's various functions were \textbf{well integrated} (Q11). P1 noted that hovering over a word in the Text Exploration View highlighted the corresponding image, effectively conveying the connection between the prompt, image, and 3D candidate set. P6 appreciated the scoring module, as it objectively reflected image quality, eliminating time-consuming subjective judgments. Finally, the majority of participants indicated that they would like to use this system \textbf{frequently} (Q12).

\textbf{Comparison of System to Baselines}

\begin{figure}
    \centering
    \includegraphics[width=1.0\linewidth]{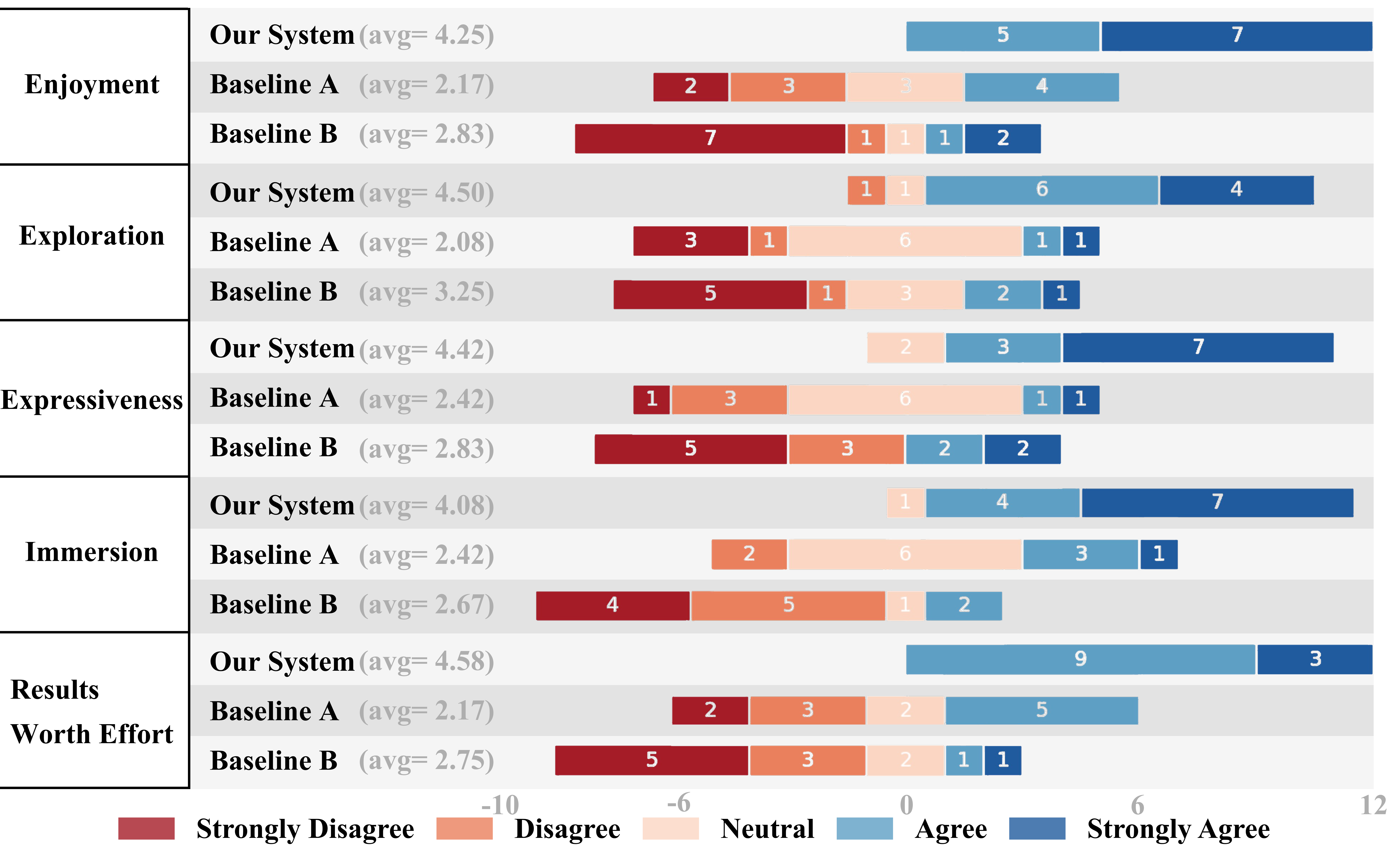}    \caption{Questionnaire results comparing our system with two baseline systems. }
    \label{fig:baseline_questionnaire}
    \vspace{-14pt}
\end{figure}

The comparative experiments employed two baselines evaluated based on the Creativity Support Index\cite{Cherry_Latulipe_2014}.
As shown in \cref{fig:baseline_questionnaire}, our system outperformed both baselines across all criteria.

Baseline A restricts users to four fixed candidates in one iteration without effective guidance for refinement options (P7), while our system enables free exploration through retrieved/generated models. P3 remarked, ``\textit{I can see a notable increase in available options, which grants me more freedom.}'' For specialized tasks like \textit{Sel3DCraft} logo design, Baseline A's 5-character text limit hindered performance (P6).

Users reported Baseline B's Image Browser contained many 3D-unsuitable images, whereas our system visually grays them out for distinction. While Baseline B provided semantic guidance, it lacked a scoring mechanism - addressed by our Text Exploration View's size-coded keywords. Additionally, some users particularly appreciated the eight-level semantic evaluation, which provided comprehensive model quality evaluation and helped them clarify their needs.

\revision{\subsubsection{Quantitative Comparison to Baselines}}
\sidecomment{R3.Q4\\SR.Q4}\revision{We also conducted quantitative comparison to baselines based on the Creativity Support Index and iteration time (see \cref{tab:summary}). The detailed paired-sample t-tests confirming statistical significance can be found in Appendix E.} 

\revision{\textbf{Our system vs. Baseline A.}
Our system enabled greater exploration freedom, with higher enjoyment and exploration scores than Baseline A ($p < .001$). It also outperformed Baseline A across all categories, offering a more immersive and rewarding experience (Immersion $p = .003$, Results worth effort $p < .001$, see Appendix E.2).}

\revision{\textbf{Our system vs. Baseline B.}
Quantitative results demonstrated that our system achieved higher levels of expressiveness ($p < .001$) and perceived value in terms of results worth the effort ($p < .001$) compared to Baseline B, highlighting its effectiveness in supporting satisfying user interactions (see Appendix E.2).}

\begin{figure}[ht]
    \centering
    \sidecomment{R1.Q8\\SR.Q3}
    \revisionbox{\includegraphics[width=0.9\linewidth]{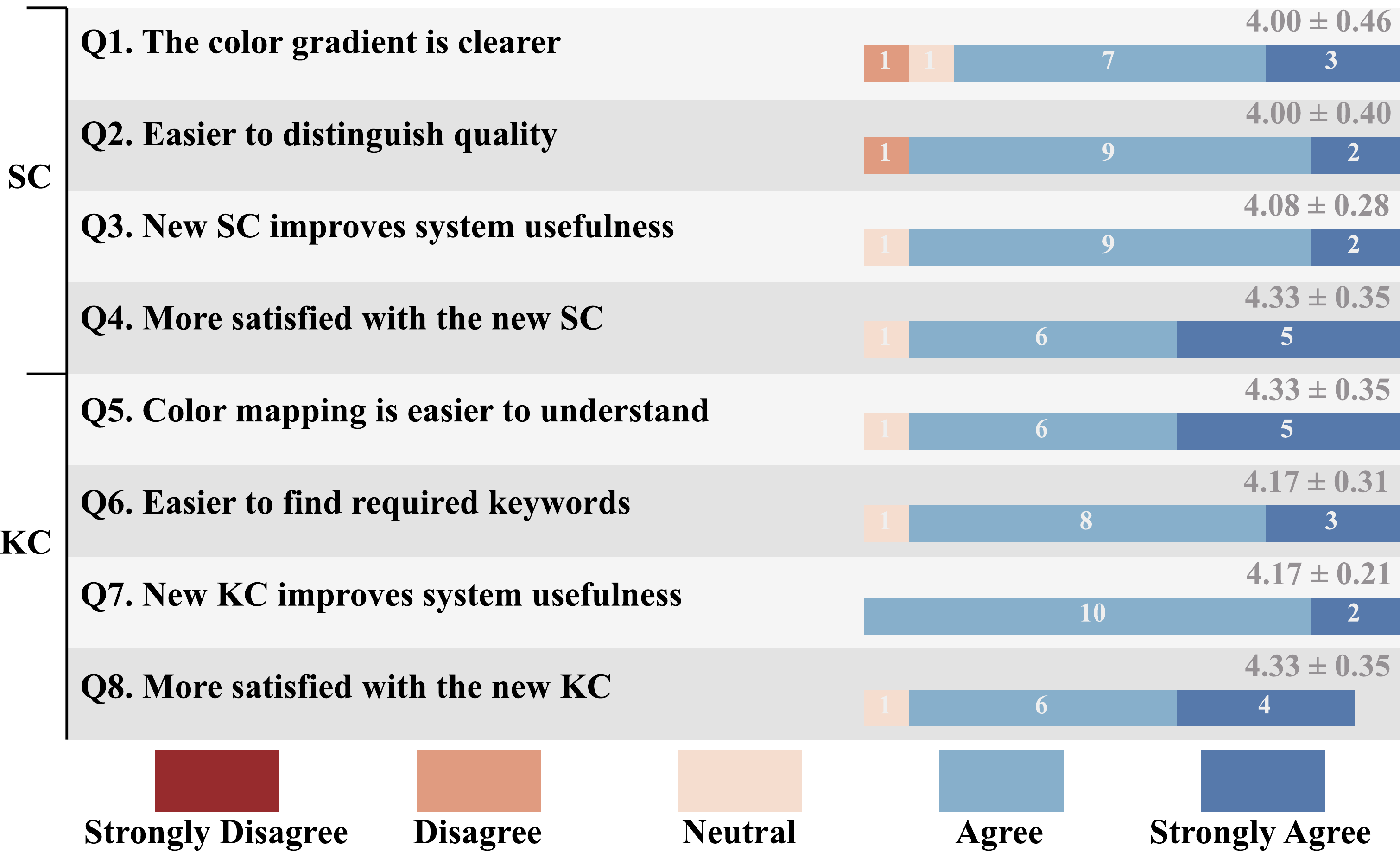}}  
    \caption{The supplementary questionnaire results on revised visualization design evaluation. SC represents the multi-view satellite chart, and KC represents the keyword contribution map. }
    \label{fig:supply_evaluation}
    \vspace{-14pt}
 \end{figure}

\begin{table}[ht]
\centering
\caption{Time analysis of the system in the user study. The proposed method shows improved efficiency in terms of total time, number of iterations, and average time per iteration compared to the baseline methods. The $\pm$ values represent the 95\% confidence intervals.}
\label{tab:summary}
\begin{tabular}{@{}lcccc@{}}
\toprule
Category & Time (s) & Iterations & Time/Iteration (s) \\ \midrule
Baseline A & 391.75 & 2.75 & 148.38  $\pm$ 40.73 \\
Baseline B & 412.58 & 3.17 & 142.40  $\pm$ 35.53 \\
Our Method & \textbf{118.83} & \textbf{1.00} & \textbf{118.83}  $\pm$ 41.73 \\ \bottomrule
\end{tabular}
\vspace{-14pt}
\end{table}
\revision{\subsection{Follow-up Evaluation of Revised Visualizations}}
\sidecomment{R1.Q8\\SR.Q3}\revision{Based on the results of our earlier evaluation (see \cref{fig:our_questionnaire}) and user feedback, we identified two specific concerns: five participants indicated that the keyword contribution map did not effectively communicate the influence of keywords on multi-view images; additionally, P3 noted that although the satellite chart was useful for assessing image quality when viewed in isolation, it caused some confusion when integrated into the full system interface. To address these issues, we refined the visual styling of both components—using distinct layouts and separable visual channels—without altering their core functionality or underlying logic (see \cref{fig:satillite,fig:keyword}).
\begin{figure}[ht]
 \centering
 \revisionbox{\includegraphics[width=0.7\linewidth]{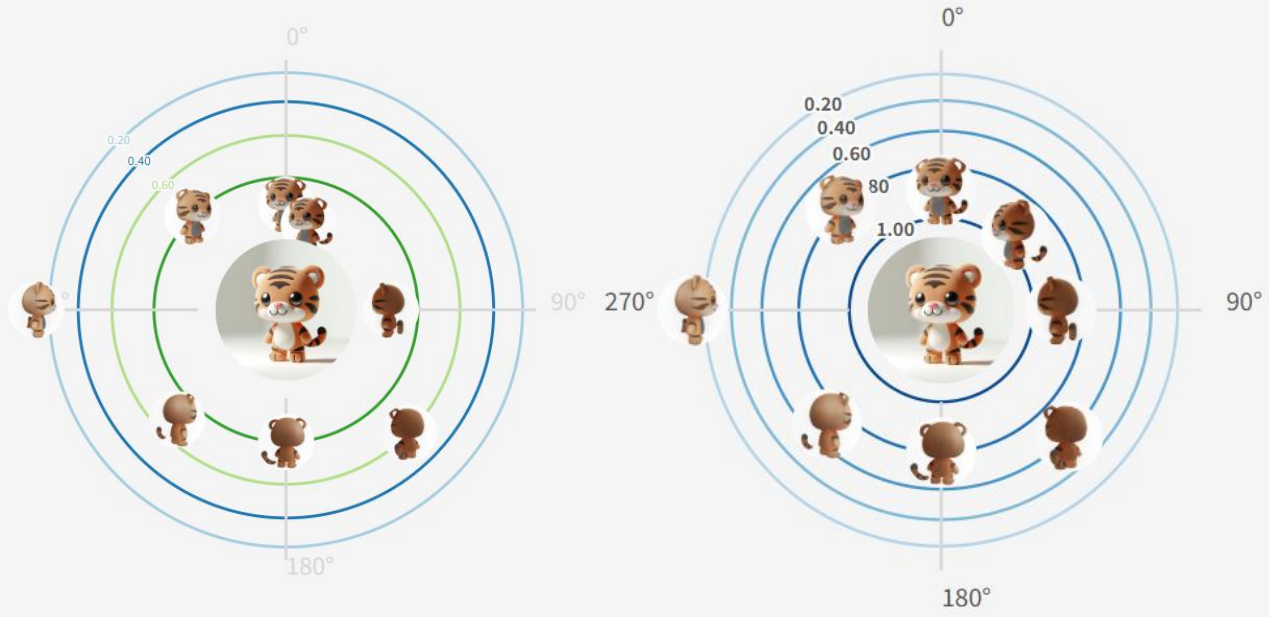}}
 \caption{
 Satellite chart before and after revision. The left shows the original version, and the right shows the revised version. }
    \label{fig:satillite}
 \vspace{-14pt}
\end{figure}

\begin{figure}[ht]
 \centering
 \revisionbox{\includegraphics[width=0.7\linewidth]{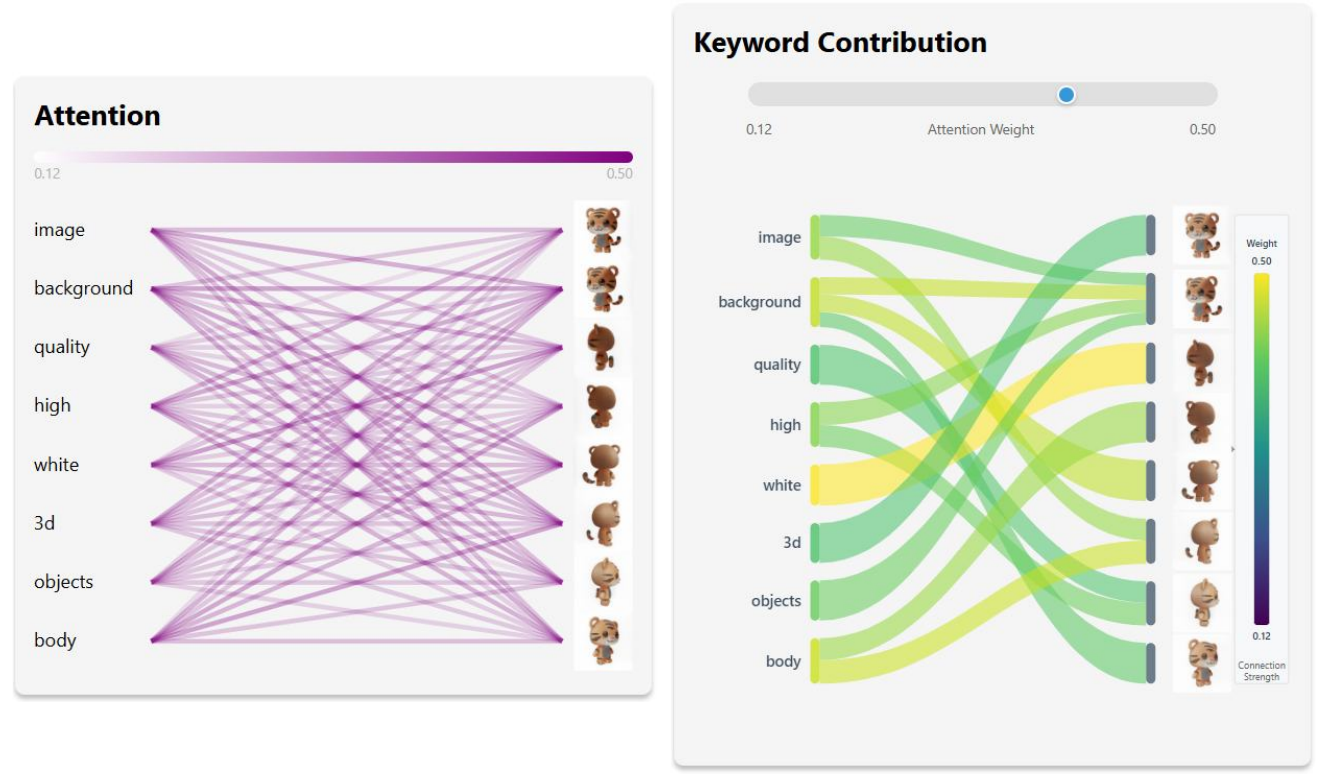}}
 \caption{
 Keyword contribution map before and after revision. The left shows the original version, and the right shows the revised version. }
    \label{fig:keyword}
 \vspace{-14pt}
\end{figure}
To further evaluate the usability of the revised visual components, we conducted a follow-up study focusing specifically on the updated multi-view satellite chart and keyword contribution map. This supplementary evaluation complements the original user study by focusing solely on the usability and clarity of the revised visual encodings. The system’s core functionality, tasks, and experimental flow remained unchanged. 
The supplementary evaluation is guided by the Principles of Effective Data Visualization~\cite{MIDWAY}, focusing on four aspects: color usage, clarity of information, user feedback, and perceived contribution to overall system usefulness.
All twelve participants from the original user study—each with over three years of experience in 3D design—were invited back for this follow-up. As they were already familiar with the system and task structure, no additional tutorial was required. Each participant selected one of the three previously defined usage scenarios and completed a design task using two versions of the system: the original and the revised.

To mitigate learning effects, the order of system usage was counterbalanced: six participants used the original system first, while the remaining six began with the revised version. Each participant was given a total of 15 minutes to use both versions. After completing the tasks, they responded to a five-point Likert-scale questionnaire. The results are summarized in \cref{fig:supply_evaluation}.}

\section{Discussion}
\subsection{Design Implications}
\sidecomment{R1.Q6}\revision{\textbf{Optimizing Keyword Distillation for Effective Prompt Refinement.} 
We score keywords and prioritize high-quality terms in the treemap wordle to balance conciseness with semantic relevance. Future work could improve keyword extraction precision.}

\revision{\textbf{Aligning Human Semantics with Machine Evaluation.} 
Our eight-level semantic evaluation framework combines computational metrics with MLLM-based assessments. User studies confirmed high satisfaction with the system's effectiveness. Future work could explore aesthetic appeal and artistic value metrics.} 

\textbf{Enhancing User-controlled Fine-tuning in Model Creation.}
When dissatisfied with results, users can either regenerate or directly modify models. For instance, in our system, some users suggested that for multi-view images with obvious defects, fine-tuning and reintroducing them into the model generation process could improve the interactive improvement of model creation. Therefore, future work should develop designer-friendly fine-tuning tools for seamless pipeline integration, or explore text-guided model adjustments -further enhancing 3D creation flexibility and human-centered interaction.

\subsection{System Generalizability}
In addition to facilitating text-to-3D prompt engineering, our system can be extended to various applications, such as game development and 3D asset management. 

For game production, it enables rapid filtering of style-consistent 3D assets across different scenarios. This contributes to improving the overall visual consistency and player experience in the game.

In 3D model library management, the system facilitates batch quality assessment for accuracy, style adherence, and diversity. This helps maintain and optimize the quality of the 3D model library while improving the efficiency of retrieval and usage.

\subsection{Limitations}
One key limitation of our system is its reliance on fixed viewpoints, which can obscure critical aspects of a 3D model. For example, in the teacup case, the lack of a top view prevents inspection of the interior, often resulting in the selection of solid, unrealistic models. This limitation can hinder users from fully evaluating geometric details and making optimal choices.

Additionally, the system does not currently support 3D animation previews, which are widely used in platforms like Sketchfab\footnote{https://sketchfab.com/}. As a result, dynamic behaviors—such as joint articulation or material deformation—cannot be assessed, limiting the system’s applicability in scenarios where animation is essential.

\sidecomment{R2.Q3\\SR.Q5}\revision{To mitigate the fixed-viewpoint limitation, we plan to augment the multi-view image evaluation stage with dedicated top-down and bird's-eye views. This will allow users to inspect occluded regions (e.g., the interior of a teacup) and thus more comprehensively assess geometric details, improving evaluation quality.

Moreover, to address the lack of animation support, we will integrate short 3D animation snippets into the final 3D viewer. Users will be able to play and navigate through animations directly within the interface, enabling assessment of dynamic behaviors (e.g., object articulation, material deformation) that static views cannot convey.}

\section{Conclusion and Future Work}
\textit{Sel3DCraft} advances text-to-3D generation through a dual-branch architecture combining model generation and retrieval. To overcome the limitations of traditional T23D tools, the system utilizes multi-view images as an intermediary, developing a multi-view hybrid scoring feature to ensure  human semantics alignment. 
The system integrates novel visualizations like treemap wordle and satellite charts for prompt recommendations and candidates visualization. We conducted user studies to validate the usefulness and effectiveness of the system. By comparing with two other baselines, we demonstrate the advantages of \textit{Sel3DCraft} in enhancing creativity in design process. 

Recent advancements in T23D have enabled the generation of 3D objects with components \cite{Chen2024ComboVerseC3, Han2024REPAROC3}. In many design scenarios, such as blind box design, designers are particularly interested in the spatial relationships of these components. Future work will focus on leveraging MLLMs for geometric analysis \cite{Abdelreheem2023ZeroShot3S}, thereby creating user-friendly exploration spaces and corresponding structured prompt engineering.



\acknowledgments{%
	The authors wish to acknowledge the support from NSSFC under Grant 22ZD05, NNSFC under Grant 62472178, and the Natural Science Foundation of Shanghai Municipality, China under Grant 24ZR1418300.%
}

\bibliographystyle{abbrv-doi-hyperref}

\bibliography{main}

\begin{thebibliography}{10}

\bibitem{Abdelreheem2023ZeroShot3S}
A.~Abdelreheem, A.~Eldesokey, M.~Ovsjanikov, and P.~Wonka.
\newblock Zero-shot 3d shape correspondence.
\newblock In J.~Kim, M.~C. Lin, and B.~Bickel, eds., {\em {SIGGRAPH} Asia 2023
  Conference Papers, {SA} 2023, Sydney, NSW, Australia, December 12-15, 2023},
  pp. 59:1--59:11. {ACM}, 2023. \href{https://doi.org/10.1145/3610548.3618228}
{doi: {{%
10\hspace{.1pt}\discretionary{.}{%
}{.}\hspace{.4pt}1145\discretionary{/}{%
}{/}3610548\hspace{.1pt}\discretionary{.}{%
}{.}\hspace{.4pt}3618228}}}


\bibitem{averkiou2014shapesynth}
M.~Averkiou, V.~G. Kim, Y.~Zheng, and N.~J. Mitra.
\newblock Shapesynth: Parameterizing model collections for coupled shape
  exploration and synthesis.
\newblock vol.~33, pp. 125--134, 2014. \href{https://doi.org/10.1111/CGF.12310}
{doi: {{%
10\hspace{.1pt}\discretionary{.}{%
}{.}\hspace{.4pt}1111\discretionary{/}{%
}{/}CGF\hspace{.1pt}\discretionary{.}{%
}{.}\hspace{.4pt}12310}}}


\bibitem{xiang2024structured}
M.~Averkiou, V.~G. Kim, Y.~Zheng, and N.~J. Mitra.
\newblock Shapesynth: Parameterizing model collections for coupled shape
  exploration and synthesis.
\newblock {\em Comput. Graph. Forum}, 33(2):125--134, 2014.
  \href{https://doi.org/10.1111/CGF.12310}
{doi: {{%
10\hspace{.1pt}\discretionary{.}{%
}{.}\hspace{.4pt}1111\discretionary{/}{%
}{/}CGF\hspace{.1pt}\discretionary{.}{%
}{.}\hspace{.4pt}12310}}}


\bibitem{bertucci2022dendromap}
D.~Bertucci, M.~M. Hamid, Y.~Anand, A.~Ruangrotsakun, D.~Tabatabai, M.~Perez,
  and M.~Kahng.
\newblock Dendromap: Visual exploration of large-scale image datasets for
  machine learning with treemaps.
\newblock {\em {IEEE} Trans. Vis. Comput. Graph.}, 29(1):320--330, 2023.
  \href{https://doi.org/10.1109/TVCG.2022.3209425}
{doi: {{%
10\hspace{.1pt}\discretionary{.}{%
}{.}\hspace{.4pt}1109\discretionary{/}{%
}{/}TVCG\hspace{.1pt}\discretionary{.}{%
}{.}\hspace{.4pt}2022\hspace{.1pt}\discretionary{.}{%
}{.}\hspace{.4pt}3209425}}}


\bibitem{brade2023promptify}
S.~Brade, B.~Wang, M.~Sousa, S.~Oore, and T.~Grossman.
\newblock Promptify: Text-to-image generation through interactive prompt
  exploration with large language models.
\newblock In S.~Follmer, J.~Han, J.~Steimle, and N.~H. Riche, eds., {\em
  Proceedings of the 36th Annual {ACM} Symposium on User Interface Software and
  Technology, {UIST} 2023, San Francisco, CA, USA, 29 October 2023- 1 November
  2023}, pp. 96:1--96:14. {ACM}, 2023.
  \href{https://doi.org/10.1145/3586183.3606725}
{doi: {{%
10\hspace{.1pt}\discretionary{.}{%
}{.}\hspace{.4pt}1145\discretionary{/}{%
}{/}3586183\hspace{.1pt}\discretionary{.}{%
}{.}\hspace{.4pt}3606725}}}


\bibitem{brown2020language}
T.~B. Brown, B.~Mann, N.~Ryder, M.~Subbiah, J.~Kaplan, P.~Dhariwal,
  A.~Neelakantan, P.~Shyam, G.~Sastry, A.~Askell, S.~Agarwal,
  A.~Herbert{-}Voss, G.~Krueger, T.~Henighan, R.~Child, A.~Ramesh, D.~M.
  Ziegler, J.~Wu, C.~Winter, C.~Hesse, M.~Chen, E.~Sigler, M.~Litwin, S.~Gray,
  B.~Chess, J.~Clark, C.~Berner, S.~McCandlish, A.~Radford, I.~Sutskever, and
  D.~Amodei.
\newblock Language models are few-shot learners.
\newblock In H.~Larochelle, M.~Ranzato, R.~Hadsell, M.~Balcan, and H.~Lin,
  eds., {\em Advances in Neural Information Processing Systems 33: Annual
  Conference on Neural Information Processing Systems 2020, NeurIPS 2020,
  December 6-12, 2020, virtual}, 2020.

\bibitem{mllmasjudge}
D.~Chen, R.~Chen, S.~Zhang, Y.~Wang, Y.~Liu, H.~Zhou, Q.~Zhang, Y.~Wan,
  P.~Zhou, and L.~Sun.
\newblock Mllm-as-a-judge: Assessing multimodal llm-as-a-judge with
  vision-language benchmark.
\newblock In {\em Forty-first International Conference on Machine Learning,
  {ICML} 2024, Vienna, Austria, July 21-27, 2024}. OpenReview.net, 2024.

\bibitem{Chen2024ComboVerseC3}
Y.~Chen, T.~Wang, T.~Wu, X.~Pan, K.~Jia, and Z.~Liu.
\newblock Comboverse: Compositional 3d assets creation using spatially-aware
  diffusion guidance.
\newblock In A.~Leonardis, E.~Ricci, S.~Roth, O.~Russakovsky, T.~Sattler, and
  G.~Varol, eds., {\em Computer Vision - {ECCV} 2024 - 18th European
  Conference, Milan, Italy, September 29-October 4, 2024, Proceedings, Part
  {XXIV}}, vol. 15082 of {\em Lecture Notes in Computer Science}, pp. 128--146.
  Springer, 2024. \href{https://doi.org/10.1007/978-3-031-72691-0_8}
{doi: {{%
10\hspace{.1pt}\discretionary{.}{%
}{.}\hspace{.4pt}1007\discretionary{/}{%
}{/}978\discretionary{%
}{-}{-}3\discretionary{%
}{-}{-}031\discretionary{%
}{-}{-}72691\discretionary{%
}{-}{-}0\_8}}}


\bibitem{Cherry_Latulipe_2014}
E.~Cherry and C.~Latulipe.
\newblock Quantifying the creativity support of digital tools through the
  creativity support index.
\newblock {\em {ACM} Trans. Comput. Hum. Interact.}, 21(4):21:1--21:25, 2014.
  \href{https://doi.org/10.1145/2617588}
{doi: {{%
10\hspace{.1pt}\discretionary{.}{%
}{.}\hspace{.4pt}1145\discretionary{/}{%
}{/}2617588}}}


\bibitem{Chung2023PromptPaintST}
J.~J.~Y. Chung and E.~Adar.
\newblock Promptpaint: Steering text-to-image generation through paint
  medium-like interactions.
\newblock pp. 6:1--6:17, 2023. \href{https://doi.org/10.1145/3586183.3606777}
{doi: {{%
10\hspace{.1pt}\discretionary{.}{%
}{.}\hspace{.4pt}1145\discretionary{/}{%
}{/}3586183\hspace{.1pt}\discretionary{.}{%
}{.}\hspace{.4pt}3606777}}}


\bibitem{2022TaleBrush}
J.~J.~Y. Chung, W.~Kim, K.~M. Yoo, H.~Lee, E.~Adar, and M.~Chang.
\newblock Talebrush: Sketching stories with generative pretrained language
  models.
\newblock In S.~D.~J. Barbosa, C.~Lampe, C.~Appert, D.~A. Shamma, S.~M.
  Drucker, J.~R. Williamson, and K.~Yatani, eds., {\em {CHI} '22: {CHI}
  Conference on Human Factors in Computing Systems, New Orleans, LA, USA, 29
  April 2022 - 5 May 2022}, pp. 209:1--209:19. {ACM}, 2022.
  \href{https://doi.org/10.1145/3491102.3501819}
{doi: {{%
10\hspace{.1pt}\discretionary{.}{%
}{.}\hspace{.4pt}1145\discretionary{/}{%
}{/}3491102\hspace{.1pt}\discretionary{.}{%
}{.}\hspace{.4pt}3501819}}}


\bibitem{objaverseXL}
M.~Deitke, R.~Liu, M.~Wallingford, H.~Ngo, O.~Michel, A.~Kusupati, A.~Fan,
  C.~Laforte, V.~Voleti, S.~Y. Gadre, E.~VanderBilt, A.~Kembhavi, C.~Vondrick,
  G.~Gkioxari, K.~Ehsani, L.~Schmidt, and A.~Farhadi.
\newblock Objaverse-xl: {A} universe of 10m+ 3d objects.
\newblock In A.~Oh, T.~Naumann, A.~Globerson, K.~Saenko, M.~Hardt, and
  S.~Levine, eds., {\em Advances in Neural Information Processing Systems 36:
  Annual Conference on Neural Information Processing Systems 2023, NeurIPS
  2023, New Orleans, LA, USA, December 10 - 16, 2023}, 2023.

\bibitem{feng2023promptmagician}
Y.~Feng, X.~Wang, K.~Wong, S.~Wang, Y.~Lu, M.~Zhu, B.~Wang, and W.~Chen.
\newblock Promptmagician: Interactive prompt engineering for text-to-image
  creation.
\newblock {\em {IEEE} Trans. Vis. Comput. Graph.}, 30(1):295--305, 2024.
  \href{https://doi.org/10.1109/TVCG.2023.3327168}
{doi: {{%
10\hspace{.1pt}\discretionary{.}{%
}{.}\hspace{.4pt}1109\discretionary{/}{%
}{/}TVCG\hspace{.1pt}\discretionary{.}{%
}{.}\hspace{.4pt}2023\hspace{.1pt}\discretionary{.}{%
}{.}\hspace{.4pt}3327168}}}


\bibitem{gao2020making}
T.~Gao, A.~Fisch, and D.~Chen.
\newblock Making pre-trained language models better few-shot learners.
\newblock In C.~Zong, F.~Xia, W.~Li, and R.~Navigli, eds., {\em Proceedings of
  the 59th Annual Meeting of the Association for Computational Linguistics and
  the 11th International Joint Conference on Natural Language Processing,
  {ACL/IJCNLP} 2021, (Volume 1: Long Papers), Virtual Event, August 1-6, 2021},
  pp. 3816--3830. Association for Computational Linguistics, 2021.
  \href{https://doi.org/10.18653/V1/2021.ACL-LONG.295}
{doi: {{%
10\hspace{.1pt}\discretionary{.}{%
}{.}\hspace{.4pt}18653\discretionary{/}{%
}{/}V1\discretionary{/}{%
}{/}2021\hspace{.1pt}\discretionary{.}{%
}{.}\hspace{.4pt}ACL\discretionary{%
}{-}{-}LONG\hspace{.1pt}\discretionary{.}{%
}{.}\hspace{.4pt}295}}}


\bibitem{Han2024REPAROC3}
H.~Han, R.~Yang, H.~Liao, J.~Xing, Z.~Xu, X.~Yu, J.~Zha, X.~Li, and W.~Li.
\newblock {REPARO:} compositional 3d assets generation with differentiable 3d
  layout alignment.
\newblock {\em CoRR}, abs/2405.18525, 2024.
  \href{https://doi.org/10.48550/ARXIV.2405.18525}
{doi: {{%
10\hspace{.1pt}\discretionary{.}{%
}{.}\hspace{.4pt}48550\discretionary{/}{%
}{/}ARXIV\hspace{.1pt}\discretionary{.}{%
}{.}\hspace{.4pt}2405\hspace{.1pt}\discretionary{.}{%
}{.}\hspace{.4pt}18525}}}


\bibitem{he2024gvgen}
X.~He, J.~Chen, S.~Peng, D.~Huang, Y.~Li, X.~Huang, C.~Yuan, W.~Ouyang, and
  T.~He.
\newblock {GVGEN:} text-to-3d generation with volumetric representation.
\newblock In A.~Leonardis, E.~Ricci, S.~Roth, O.~Russakovsky, T.~Sattler, and
  G.~Varol, eds., {\em Computer Vision - {ECCV} 2024 - 18th European
  Conference, Milan, Italy, September 29-October 4, 2024, Proceedings, Part
  {VIII}}, vol. 15066 of {\em Lecture Notes in Computer Science}, pp. 463--479.
  Springer, 2024. \href{https://doi.org/10.1007/978-3-031-73242-3_26}
{doi: {{%
10\hspace{.1pt}\discretionary{.}{%
}{.}\hspace{.4pt}1007\discretionary{/}{%
}{/}978\discretionary{%
}{-}{-}3\discretionary{%
}{-}{-}031\discretionary{%
}{-}{-}73242\discretionary{%
}{-}{-}3\_26}}}


\bibitem{hessel2021clipscore}
J.~Hessel, A.~Holtzman, M.~Forbes, R.~L. Bras, and Y.~Choi.
\newblock Clipscore: {A} reference-free evaluation metric for image captioning.
\newblock In M.~Moens, X.~Huang, L.~Specia, and S.~W. Yih, eds., {\em
  Proceedings of the 2021 Conference on Empirical Methods in Natural Language
  Processing, {EMNLP} 2021, Virtual Event / Punta Cana, Dominican Republic,
  7-11 November, 2021}, pp. 7514--7528. Association for Computational
  Linguistics, 2021. \href{https://doi.org/10.18653/V1/2021.EMNLP-MAIN.595}
{doi: {{%
10\hspace{.1pt}\discretionary{.}{%
}{.}\hspace{.4pt}18653\discretionary{/}{%
}{/}V1\discretionary{/}{%
}{/}2021\hspace{.1pt}\discretionary{.}{%
}{.}\hspace{.4pt}EMNLP\discretionary{%
}{-}{-}MAIN\hspace{.1pt}\discretionary{.}{%
}{.}\hspace{.4pt}595}}}


\bibitem{hong2023lrm}
Y.~Hong, K.~Zhang, J.~Gu, S.~Bi, Y.~Zhou, D.~Liu, F.~Liu, K.~Sunkavalli,
  T.~Bui, and H.~Tan.
\newblock {LRM:} large reconstruction model for single image to 3d.
\newblock In {\em The Twelfth International Conference on Learning
  Representations, {ICLR} 2024, Vienna, Austria, May 7-11, 2024}.
  OpenReview.net, 2024.

\bibitem{hou2007saliency}
X.~Hou and L.~Zhang.
\newblock Saliency detection: {A} spectral residual approach.
\newblock In {\em 2007 {IEEE} Computer Society Conference on Computer Vision
  and Pattern Recognition {(CVPR} 2007), 18-23 June 2007, Minneapolis,
  Minnesota, {USA}}. {IEEE} Computer Society, 2007.
  \href{https://doi.org/10.1109/CVPR.2007.383267}
{doi: {{%
10\hspace{.1pt}\discretionary{.}{%
}{.}\hspace{.4pt}1109\discretionary{/}{%
}{/}CVPR\hspace{.1pt}\discretionary{.}{%
}{.}\hspace{.4pt}2007\hspace{.1pt}\discretionary{.}{%
}{.}\hspace{.4pt}383267}}}


\bibitem{wong_anchorage_2023}
W.~Kam{-}Kwai, X.~Wang, Y.~Wang, J.~He, R.~Zhang, and H.~Qu.
\newblock Anchorage: Visual analysis of satisfaction in customer service videos
  via anchor events.
\newblock {\em {IEEE} Trans. Vis. Comput. Graph.}, 30(7):4008--4022, 2024.
  \href{https://doi.org/10.1109/TVCG.2023.3245609}
{doi: {{%
10\hspace{.1pt}\discretionary{.}{%
}{.}\hspace{.4pt}1109\discretionary{/}{%
}{/}TVCG\hspace{.1pt}\discretionary{.}{%
}{.}\hspace{.4pt}2023\hspace{.1pt}\discretionary{.}{%
}{.}\hspace{.4pt}3245609}}}


\bibitem{kirillov2023segany}
A.~Kirillov, E.~Mintun, N.~Ravi, H.~Mao, C.~Rolland, L.~Gustafson, T.~Xiao,
  S.~Whitehead, A.~C. Berg, W.~Lo, P.~Doll{\'{a}}r, and R.~B. Girshick.
\newblock Segment anything.
\newblock In {\em {IEEE/CVF} International Conference on Computer Vision,
  {ICCV} 2023, Paris, France, October 1-6, 2023}, pp. 3992--4003. {IEEE}, 2023.
  \href{https://doi.org/10.1109/ICCV51070.2023.00371}
{doi: {{%
10\hspace{.1pt}\discretionary{.}{%
}{.}\hspace{.4pt}1109\discretionary{/}{%
}{/}ICCV51070\hspace{.1pt}\discretionary{.}{%
}{.}\hspace{.4pt}2023\hspace{.1pt}\discretionary{.}{%
}{.}\hspace{.4pt}00371}}}


\bibitem{li2022blip}
J.~Li, D.~Li, C.~Xiong, and S.~C.~H. Hoi.
\newblock {BLIP:} bootstrapping language-image pre-training for unified
  vision-language understanding and generation.
\newblock In K.~Chaudhuri, S.~Jegelka, L.~Song, C.~Szepesv{\'{a}}ri, G.~Niu,
  and S.~Sabato, eds., {\em International Conference on Machine Learning,
  {ICML} 2022, 17-23 July 2022, Baltimore, Maryland, {USA}}, vol. 162 of {\em
  Proceedings of Machine Learning Research}, pp. 12888--12900. {PMLR}, 2022.

\bibitem{li2023instant3d}
J.~Li, H.~Tan, K.~Zhang, Z.~Xu, F.~Luan, Y.~Xu, Y.~Hong, K.~Sunkavalli,
  G.~Shakhnarovich, and S.~Bi.
\newblock Instant3d: Fast text-to-3d with sparse-view generation and large
  reconstruction model.
\newblock In {\em The Twelfth International Conference on Learning
  Representations, {ICLR} 2024, Vienna, Austria, May 7-11, 2024}.
  OpenReview.net, 2024.

\bibitem{10.1109/TVCG.2024.3388514}
Y.~Li, J.~Wang, P.~Aboagye, C.-C.~M. Yeh, Y.~Zheng, L.~Wang, W.~Zhang, and
  K.-L. Ma.
\newblock Visual analytics for efficient image exploration and user-guided
  image captioning.
\newblock {\em IEEE Transactions on Visualization and Computer Graphics},
  30(6):2875–2887,  13 pages, apr 2024.
  \href{https://doi.org/10.1109/TVCG.2024.3388514}
{doi: {{%
10\hspace{.1pt}\discretionary{.}{%
}{.}\hspace{.4pt}1109\discretionary{/}{%
}{/}TVCG\hspace{.1pt}\discretionary{.}{%
}{.}\hspace{.4pt}2024\hspace{.1pt}\discretionary{.}{%
}{.}\hspace{.4pt}3388514}}}


\bibitem{liang2022multiviz}
P.~P. Liang, Y.~Lyu, G.~Chhablani, N.~Jain, Z.~Deng, X.~Wang, L.~Morency, and
  R.~Salakhutdinov.
\newblock Multiviz: Towards visualizing and understanding multimodal models.
\newblock In {\em The Eleventh International Conference on Learning
  Representations, {ICLR} 2023, Kigali, Rwanda, May 1-5, 2023}. OpenReview.net,
  2023.

\bibitem{liu2023openshape}
M.~Liu, R.~Shi, K.~Kuang, Y.~Zhu, X.~Li, S.~Han, H.~Cai, F.~Porikli, and H.~Su.
\newblock Openshape: Scaling up 3d shape representation towards open-world
  understanding.
\newblock In A.~Oh, T.~Naumann, A.~Globerson, K.~Saenko, M.~Hardt, and
  S.~Levine, eds., {\em Advances in Neural Information Processing Systems 36:
  Annual Conference on Neural Information Processing Systems 2023, NeurIPS
  2023, New Orleans, LA, USA, December 10 - 16, 2023}, 2023.

\bibitem{liu2023one}
M.~Liu, C.~Xu, H.~Jin, L.~Chen, M.~V. T., Z.~Xu, and H.~Su.
\newblock One-2-3-45: Any single image to 3d mesh in 45 seconds without
  per-shape optimization.
\newblock In A.~Oh, T.~Naumann, A.~Globerson, K.~Saenko, M.~Hardt, and
  S.~Levine, eds., {\em Advances in Neural Information Processing Systems 36:
  Annual Conference on Neural Information Processing Systems 2023, NeurIPS
  2023, New Orleans, LA, USA, December 10 - 16, 2023}, 2023.

\bibitem{liu2023zero}
R.~Liu, R.~Wu, B.~V. Hoorick, P.~Tokmakov, S.~Zakharov, and C.~Vondrick.
\newblock Zero-1-to-3: Zero-shot one image to 3d object.
\newblock In {\em {IEEE/CVF} International Conference on Computer Vision,
  {ICCV} 2023, Paris, France, October 1-6, 2023}, pp. 9264--9275. {IEEE}, 2023.
  \href{https://doi.org/10.1109/ICCV51070.2023.00853}
{doi: {{%
10\hspace{.1pt}\discretionary{.}{%
}{.}\hspace{.4pt}1109\discretionary{/}{%
}{/}ICCV51070\hspace{.1pt}\discretionary{.}{%
}{.}\hspace{.4pt}2023\hspace{.1pt}\discretionary{.}{%
}{.}\hspace{.4pt}00853}}}


\bibitem{liu2022design}
V.~Liu and L.~B. Chilton.
\newblock Design guidelines for prompt engineering text-to-image generative
  models.
\newblock In S.~D.~J. Barbosa, C.~Lampe, C.~Appert, D.~A. Shamma, S.~M.
  Drucker, J.~R. Williamson, and K.~Yatani, eds., {\em {CHI} '22: {CHI}
  Conference on Human Factors in Computing Systems, New Orleans, LA, USA, 29
  April 2022 - 5 May 2022}, pp. 384:1--384:23. {ACM}, 2022.
  \href{https://doi.org/10.1145/3491102.3501825}
{doi: {{%
10\hspace{.1pt}\discretionary{.}{%
}{.}\hspace{.4pt}1145\discretionary{/}{%
}{/}3491102\hspace{.1pt}\discretionary{.}{%
}{.}\hspace{.4pt}3501825}}}


\bibitem{Liu20223DALLEIT}
V.~Liu, J.~Vermeulen, G.~W. Fitzmaurice, and J.~Matejka.
\newblock 3dall-e: Integrating text-to-image {AI} in 3d design workflows.
\newblock In D.~Byrne, N.~Martelaro, A.~Boucher, D.~J. Chatting, S.~F. Alaoui,
  S.~E. Fox, I.~Nicenboim, and C.~MacArthur, eds., {\em Proceedings of the 2023
  {ACM} Designing Interactive Systems Conference, {DIS} 2023, Pittsburgh, PA,
  USA, July 10-14, 2023}, pp. 1955--1977. {ACM}, 2023.
  \href{https://doi.org/10.1145/3563657.3596098}
{doi: {{%
10\hspace{.1pt}\discretionary{.}{%
}{.}\hspace{.4pt}1145\discretionary{/}{%
}{/}3563657\hspace{.1pt}\discretionary{.}{%
}{.}\hspace{.4pt}3596098}}}


\bibitem{long2023wonder3d}
X.~Long, Y.~Guo, C.~Lin, Y.~Liu, Z.~Dou, L.~Liu, Y.~Ma, S.~Zhang, M.~Habermann,
  C.~Theobalt, and W.~Wang.
\newblock Wonder3d: Single image to 3d using cross-domain diffusion.
\newblock In {\em {IEEE/CVF} Conference on Computer Vision and Pattern
  Recognition, {CVPR} 2024, Seattle, WA, USA, June 16-22, 2024}, pp.
  9970--9980. {IEEE}, 2024. \href{https://doi.org/10.1109/CVPR52733.2024.00951}
{doi: {{%
10\hspace{.1pt}\discretionary{.}{%
}{.}\hspace{.4pt}1109\discretionary{/}{%
}{/}CVPR52733\hspace{.1pt}\discretionary{.}{%
}{.}\hspace{.4pt}2024\hspace{.1pt}\discretionary{.}{%
}{.}\hspace{.4pt}00951}}}


\bibitem{marks1997design}
J.~Marks, B.~Andalman, P.~A. Beardsley, W.~T. Freeman, S.~F. Gibson, J.~K.
  Hodgins, T.~Kang, B.~Mirtich, H.~Pfister, W.~Ruml, K.~Ryall, J.~E. Seims, and
  S.~M. Shieber.
\newblock Design galleries: a general approach to setting parameters for
  computer graphics and animation.
\newblock In G.~S. Owen, T.~Whitted, and B.~Mones{-}Hattal, eds., {\em
  Proceedings of the 24th Annual Conference on Computer Graphics and
  Interactive Techniques, {SIGGRAPH} 1997, Los Angeles, CA, USA, August 3-8,
  1997}, pp. 389--400. {ACM}, 1997.
  \href{https://doi.org/10.1145/258734.258887}
{doi: {{%
10\hspace{.1pt}\discretionary{.}{%
}{.}\hspace{.4pt}1145\discretionary{/}{%
}{/}258734\hspace{.1pt}\discretionary{.}{%
}{.}\hspace{.4pt}258887}}}


\bibitem{MIDWAY}
S.~R. Midway.
\newblock Principles of effective data visualization.
\newblock {\em Patterns}, 1(9):100141, 2020.
  \href{https://doi.org/10.1016/j.patter.2020.100141}
{doi: {{%
10\hspace{.1pt}\discretionary{.}{%
}{.}\hspace{.4pt}1016\discretionary{/}{%
}{/}j\hspace{.1pt}\discretionary{.}{%
}{.}\hspace{.4pt}patter\hspace{.1pt}\discretionary{.}{%
}{.}\hspace{.4pt}2020\hspace{.1pt}\discretionary{.}{%
}{.}\hspace{.4pt}100141}}}


\bibitem{mildenhall2020nerf}
B.~Mildenhall, P.~P. Srinivasan, M.~Tancik, J.~T. Barron, R.~Ramamoorthi, and
  R.~Ng.
\newblock Nerf: Representing scenes as neural radiance fields for view
  synthesis.
\newblock In A.~Vedaldi, H.~Bischof, T.~Brox, and J.~Frahm, eds., {\em Computer
  Vision - {ECCV} 2020 - 16th European Conference, Glasgow, UK, August 23-28,
  2020, Proceedings, Part {I}}, vol. 12346 of {\em Lecture Notes in Computer
  Science}, pp. 405--421. Springer, 2020.
  \href{https://doi.org/10.1007/978-3-030-58452-8_24}
{doi: {{%
10\hspace{.1pt}\discretionary{.}{%
}{.}\hspace{.4pt}1007\discretionary{/}{%
}{/}978\discretionary{%
}{-}{-}3\discretionary{%
}{-}{-}030\discretionary{%
}{-}{-}58452\discretionary{%
}{-}{-}8\_24}}}


\bibitem{nichol2022point}
A.~Nichol, H.~Jun, P.~Dhariwal, P.~Mishkin, and M.~Chen.
\newblock Point-e: {A} system for generating 3d point clouds from complex
  prompts.
\newblock {\em CoRR}, abs/2212.08751, 2022.
  \href{https://doi.org/10.48550/ARXIV.2212.08751}
{doi: {{%
10\hspace{.1pt}\discretionary{.}{%
}{.}\hspace{.4pt}48550\discretionary{/}{%
}{/}ARXIV\hspace{.1pt}\discretionary{.}{%
}{.}\hspace{.4pt}2212\hspace{.1pt}\discretionary{.}{%
}{.}\hspace{.4pt}08751}}}


\bibitem{nichol2021glide}
A.~Q. Nichol, P.~Dhariwal, A.~Ramesh, P.~Shyam, P.~Mishkin, B.~McGrew,
  I.~Sutskever, and M.~Chen.
\newblock {GLIDE:} towards photorealistic image generation and editing with
  text-guided diffusion models.
\newblock In K.~Chaudhuri, S.~Jegelka, L.~Song, C.~Szepesv{\'{a}}ri, G.~Niu,
  and S.~Sabato, eds., {\em International Conference on Machine Learning,
  {ICML} 2022, 17-23 July 2022, Baltimore, Maryland, {USA}}, vol. 162 of {\em
  Proceedings of Machine Learning Research}, pp. 16784--16804. {PMLR}, 2022.

\bibitem{oppenlaender2022prompt}
J.~Oppenlaender.
\newblock Prompt engineering for text-based generative art.
\newblock {\em CoRR}, abs/2204.13988, 2022.
  \href{https://doi.org/10.48550/ARXIV.2204.13988}
{doi: {{%
10\hspace{.1pt}\discretionary{.}{%
}{.}\hspace{.4pt}48550\discretionary{/}{%
}{/}ARXIV\hspace{.1pt}\discretionary{.}{%
}{.}\hspace{.4pt}2204\hspace{.1pt}\discretionary{.}{%
}{.}\hspace{.4pt}13988}}}


\bibitem{ouyang2022training}
L.~Ouyang, J.~Wu, X.~Jiang, D.~Almeida, C.~L. Wainwright, P.~Mishkin, C.~Zhang,
  S.~Agarwal, K.~Slama, A.~Ray, J.~Schulman, J.~Hilton, F.~Kelton, L.~Miller,
  M.~Simens, A.~Askell, P.~Welinder, P.~F. Christiano, J.~Leike, and R.~Lowe.
\newblock Training language models to follow instructions with human feedback.
\newblock 2022.

\bibitem{pandey2023juxtaform}
K.~Pandey, F.~Chevalier, and K.~Singh.
\newblock Juxtaform: interactive visual summarization for exploratory shape
  design.
\newblock {\em {ACM} Trans. Graph.}, 42(4):52:1--52:14, 2023.
  \href{https://doi.org/10.1145/3592436}
{doi: {{%
10\hspace{.1pt}\discretionary{.}{%
}{.}\hspace{.4pt}1145\discretionary{/}{%
}{/}3592436}}}


\bibitem{podell2023sdxl}
D.~Podell, Z.~English, K.~Lacey, A.~Blattmann, T.~Dockhorn, J.~M{\"{u}}ller,
  J.~Penna, and R.~Rombach.
\newblock {SDXL:} improving latent diffusion models for high-resolution image
  synthesis.
\newblock 2024.

\bibitem{poole2022dreamfusion}
B.~Poole, A.~Jain, J.~T. Barron, and B.~Mildenhall.
\newblock Dreamfusion: Text-to-3d using 2d diffusion.
\newblock In {\em The Eleventh International Conference on Learning
  Representations, {ICLR} 2023, Kigali, Rwanda, May 1-5, 2023}. OpenReview.net,
  2023.

\bibitem{radford2021CLIP}
A.~Radford, J.~W. Kim, C.~Hallacy, A.~Ramesh, G.~Goh, S.~Agarwal, G.~Sastry,
  A.~Askell, P.~Mishkin, J.~Clark, G.~Krueger, and I.~Sutskever.
\newblock Learning transferable visual models from natural language
  supervision.
\newblock In M.~Meila and T.~Zhang, eds., {\em Proceedings of the 38th
  International Conference on Machine Learning, {ICML} 2021, 18-24 July 2021,
  Virtual Event}, vol. 139 of {\em Proceedings of Machine Learning Research},
  pp. 8748--8763. {PMLR}, 2021.

\bibitem{radford2021learning}
A.~Radford, J.~W. Kim, C.~Hallacy, A.~Ramesh, G.~Goh, S.~Agarwal, G.~Sastry,
  A.~Askell, P.~Mishkin, J.~Clark, G.~Krueger, and I.~Sutskever.
\newblock Learning transferable visual models from natural language
  supervision.
\newblock In M.~Meila and T.~Zhang, eds., {\em Proceedings of the 38th
  International Conference on Machine Learning, {ICML} 2021, 18-24 July 2021,
  Virtual Event}, vol. 139 of {\em Proceedings of Machine Learning Research},
  pp. 8748--8763. {PMLR}, 2021.

\bibitem{ramesh2022hierarchical}
A.~Ramesh, P.~Dhariwal, A.~Nichol, C.~Chu, and M.~Chen.
\newblock Hierarchical text-conditional image generation with {CLIP} latents.
\newblock {\em CoRR}, abs/2204.06125, 2022.
  \href{https://doi.org/10.48550/ARXIV.2204.06125}
{doi: {{%
10\hspace{.1pt}\discretionary{.}{%
}{.}\hspace{.4pt}48550\discretionary{/}{%
}{/}ARXIV\hspace{.1pt}\discretionary{.}{%
}{.}\hspace{.4pt}2204\hspace{.1pt}\discretionary{.}{%
}{.}\hspace{.4pt}06125}}}


\bibitem{rombach2022high}
R.~Rombach, A.~Blattmann, D.~Lorenz, P.~Esser, and B.~Ommer.
\newblock High-resolution image synthesis with latent diffusion models.
\newblock In {\em {IEEE/CVF} Conference on Computer Vision and Pattern
  Recognition, {CVPR} 2022, New Orleans, LA, USA, June 18-24, 2022}, pp.
  10674--10685. {IEEE}, 2022.
  \href{https://doi.org/10.1109/CVPR52688.2022.01042}
{doi: {{%
10\hspace{.1pt}\discretionary{.}{%
}{.}\hspace{.4pt}1109\discretionary{/}{%
}{/}CVPR52688\hspace{.1pt}\discretionary{.}{%
}{.}\hspace{.4pt}2022\hspace{.1pt}\discretionary{.}{%
}{.}\hspace{.4pt}01042}}}


\bibitem{shi2023mvdream}
Y.~Shi, P.~Wang, J.~Ye, L.~Mai, K.~Li, and X.~Yang.
\newblock Mvdream: Multi-view diffusion for 3d generation.
\newblock In {\em The Twelfth International Conference on Learning
  Representations, {ICLR} 2024, Vienna, Austria, May 7-11, 2024}.
  OpenReview.net, 2024.

\bibitem{shin2020autoprompt}
T.~Shin, Y.~Razeghi, R.~L.~L. IV, E.~Wallace, and S.~Singh.
\newblock Autoprompt: Eliciting knowledge from language models with
  automatically generated prompts.
\newblock In B.~Webber, T.~Cohn, Y.~He, and Y.~Liu, eds., {\em Proceedings of
  the 2020 Conference on Empirical Methods in Natural Language Processing,
  {EMNLP} 2020, Online, November 16-20, 2020}, pp. 4222--4235. Association for
  Computational Linguistics, 2020.
  \href{https://doi.org/10.18653/V1/2020.EMNLP-MAIN.346}
{doi: {{%
10\hspace{.1pt}\discretionary{.}{%
}{.}\hspace{.4pt}18653\discretionary{/}{%
}{/}V1\discretionary{/}{%
}{/}2020\hspace{.1pt}\discretionary{.}{%
}{.}\hspace{.4pt}EMNLP\discretionary{%
}{-}{-}MAIN\hspace{.1pt}\discretionary{.}{%
}{.}\hspace{.4pt}346}}}


\bibitem{Son2023GenQuerySE}
K.~Son, D.~Choi, T.~S. Kim, Y.~Kim, and J.~Kim.
\newblock Genquery: Supporting expressive visual search with generative models.
\newblock In F.~F. Mueller, P.~Kyburz, J.~R. Williamson, C.~Sas, M.~L. Wilson,
  P.~O.~T. Dugas, and I.~Shklovski, eds., {\em Proceedings of the {CHI}
  Conference on Human Factors in Computing Systems, {CHI} 2024, Honolulu, HI,
  USA, May 11-16, 2024}, pp. 180:1--180:19. {ACM}, 2024.
  \href{https://doi.org/10.1145/3613904.3642847}
{doi: {{%
10\hspace{.1pt}\discretionary{.}{%
}{.}\hspace{.4pt}1145\discretionary{/}{%
}{/}3613904\hspace{.1pt}\discretionary{.}{%
}{.}\hspace{.4pt}3642847}}}


\bibitem{strobelt2022interactive}
H.~Strobelt, A.~Webson, V.~Sanh, B.~Hoover, J.~Beyer, H.~Pfister, and A.~M.
  Rush.
\newblock Interactive and visual prompt engineering for ad-hoc task adaptation
  with large language models.
\newblock {\em {IEEE} Trans. Vis. Comput. Graph.}, 29(1):1146--1156, 2023.
  \href{https://doi.org/10.1109/TVCG.2022.3209479}
{doi: {{%
10\hspace{.1pt}\discretionary{.}{%
}{.}\hspace{.4pt}1109\discretionary{/}{%
}{/}TVCG\hspace{.1pt}\discretionary{.}{%
}{.}\hspace{.4pt}2022\hspace{.1pt}\discretionary{.}{%
}{.}\hspace{.4pt}3209479}}}


\bibitem{Swearngin2020ScoutRE}
A.~Swearngin, C.~Wang, A.~Oleson, J.~Fogarty, and A.~J. Ko.
\newblock Scout: Rapid exploration of interface layout alternatives through
  high-level design constraints.
\newblock In R.~Bernhaupt, F.~F. Mueller, D.~Verweij, J.~Andres, J.~McGrenere,
  A.~Cockburn, I.~Avellino, A.~Goguey, P.~Bj{\o}n, S.~Zhao, B.~P. Samson, and
  R.~Kocielnik, eds., {\em {CHI} '20: {CHI} Conference on Human Factors in
  Computing Systems, Honolulu, HI, USA, April 25-30, 2020}, pp. 1--13. {ACM},
  2020. \href{https://doi.org/10.1145/3313831.3376593}
{doi: {{%
10\hspace{.1pt}\discretionary{.}{%
}{.}\hspace{.4pt}1145\discretionary{/}{%
}{/}3313831\hspace{.1pt}\discretionary{.}{%
}{.}\hspace{.4pt}3376593}}}


\bibitem{tang2024lgm}
J.~Tang, Z.~Chen, X.~Chen, T.~Wang, G.~Zeng, and Z.~Liu.
\newblock {LGM:} large multi-view gaussian model for high-resolution 3d content
  creation.
\newblock In A.~Leonardis, E.~Ricci, S.~Roth, O.~Russakovsky, T.~Sattler, and
  G.~Varol, eds., {\em Computer Vision - {ECCV} 2024 - 18th European
  Conference, Milan, Italy, September 29-October 4, 2024, Proceedings, Part
  {IV}}, vol. 15062 of {\em Lecture Notes in Computer Science}, pp. 1--18.
  Springer, 2024. \href{https://doi.org/10.1007/978-3-031-73235-5_1}
{doi: {{%
10\hspace{.1pt}\discretionary{.}{%
}{.}\hspace{.4pt}1007\discretionary{/}{%
}{/}978\discretionary{%
}{-}{-}3\discretionary{%
}{-}{-}031\discretionary{%
}{-}{-}73235\discretionary{%
}{-}{-}5\_1}}}


\bibitem{tang2023dreamgaussian}
J.~Tang, J.~Ren, H.~Zhou, Z.~Liu, and G.~Zeng.
\newblock Dreamgaussian: Generative gaussian splatting for efficient 3d content
  creation.
\newblock In {\em The Twelfth International Conference on Learning
  Representations, {ICLR} 2024, Vienna, Austria, May 7-11, 2024}.
  OpenReview.net, 2024.

\bibitem{tang2023volumediffusion}
Z.~Tang, S.~Gu, C.~Wang, T.~Zhang, J.~Bao, D.~Chen, and B.~Guo.
\newblock Volumediffusion: Flexible text-to-3d generation with efficient
  volumetric encoder.
\newblock {\em CoRR}, abs/2312.11459, 2023.
  \href{https://doi.org/10.48550/ARXIV.2312.11459}
{doi: {{%
10\hspace{.1pt}\discretionary{.}{%
}{.}\hspace{.4pt}48550\discretionary{/}{%
}{/}ARXIV\hspace{.1pt}\discretionary{.}{%
}{.}\hspace{.4pt}2312\hspace{.1pt}\discretionary{.}{%
}{.}\hspace{.4pt}11459}}}


\bibitem{kailath1967divergence}
G.~T. Toussaint.
\newblock Comments on "the divergence and bhattacharyya distance measures in
  signal selection".
\newblock {\em {IEEE} Trans. Commun.}, 20(3):485, 1972.
  \href{https://doi.org/10.1109/TCOM.1972.1091157}
{doi: {{%
10\hspace{.1pt}\discretionary{.}{%
}{.}\hspace{.4pt}1109\discretionary{/}{%
}{/}TCOM\hspace{.1pt}\discretionary{.}{%
}{.}\hspace{.4pt}1972\hspace{.1pt}\discretionary{.}{%
}{.}\hspace{.4pt}1091157}}}


\bibitem{wang2023democratizing}
B.~Wang.
\newblock Democratizing content creation and consumption through human-ai
  copilot systems.
\newblock In S.~Follmer, J.~Han, J.~Steimle, and N.~H. Riche, eds., {\em
  Adjunct Proceedings of the 36th Annual {ACM} Symposium on User Interface
  Software and Technology, {UIST} 2023, San Francisco, CA, USA, 29 October
  2023- 1 November 2023}, pp. 105:1--105:4. {ACM}, 2023.
  \href{https://doi.org/10.1145/3586182.3616707}
{doi: {{%
10\hspace{.1pt}\discretionary{.}{%
}{.}\hspace{.4pt}1145\discretionary{/}{%
}{/}3586182\hspace{.1pt}\discretionary{.}{%
}{.}\hspace{.4pt}3616707}}}


\bibitem{wang2023score}
H.~Wang, X.~Du, J.~Li, R.~A. Yeh, and G.~Shakhnarovich.
\newblock Score jacobian chaining: Lifting pretrained 2d diffusion models for
  3d generation.
\newblock In {\em {IEEE/CVF} Conference on Computer Vision and Pattern
  Recognition, {CVPR} 2023, Vancouver, BC, Canada, June 17-24, 2023}, pp.
  12619--12629. {IEEE}, 2023.
  \href{https://doi.org/10.1109/CVPR52729.2023.01214}
{doi: {{%
10\hspace{.1pt}\discretionary{.}{%
}{.}\hspace{.4pt}1109\discretionary{/}{%
}{/}CVPR52729\hspace{.1pt}\discretionary{.}{%
}{.}\hspace{.4pt}2023\hspace{.1pt}\discretionary{.}{%
}{.}\hspace{.4pt}01214}}}


\bibitem{wang2021m2lens}
X.~Wang, J.~He, Z.~Jin, M.~Yang, Y.~Wang, and H.~Qu.
\newblock M2lens: Visualizing and explaining multimodal models for sentiment
  analysis.
\newblock {\em {IEEE} Trans. Vis. Comput. Graph.}, 28(1):802--812, 2022.
  \href{https://doi.org/10.1109/TVCG.2021.3114794}
{doi: {{%
10\hspace{.1pt}\discretionary{.}{%
}{.}\hspace{.4pt}1109\discretionary{/}{%
}{/}TVCG\hspace{.1pt}\discretionary{.}{%
}{.}\hspace{.4pt}2021\hspace{.1pt}\discretionary{.}{%
}{.}\hspace{.4pt}3114794}}}


\bibitem{wang2022diffusiondb}
Z.~J. Wang, E.~Montoya, D.~Munechika, H.~Yang, B.~Hoover, and D.~H. Chau.
\newblock Diffusiondb: {A} large-scale prompt gallery dataset for text-to-image
  generative models.
\newblock In A.~Rogers, J.~L. Boyd{-}Graber, and N.~Okazaki, eds., {\em
  Proceedings of the 61st Annual Meeting of the Association for Computational
  Linguistics (Volume 1: Long Papers), {ACL} 2023, Toronto, Canada, July 9-14,
  2023}, pp. 893--911. Association for Computational Linguistics, 2023.
  \href{https://doi.org/10.18653/V1/2023.ACL-LONG.51}
{doi: {{%
10\hspace{.1pt}\discretionary{.}{%
}{.}\hspace{.4pt}18653\discretionary{/}{%
}{/}V1\discretionary{/}{%
}{/}2023\hspace{.1pt}\discretionary{.}{%
}{.}\hspace{.4pt}ACL\discretionary{%
}{-}{-}LONG\hspace{.1pt}\discretionary{.}{%
}{.}\hspace{.4pt}51}}}


\bibitem{wu2022promptchainer}
T.~Wu, E.~Jiang, A.~Donsbach, J.~Gray, A.~Molina, M.~Terry, and C.~J. Cai.
\newblock Promptchainer: Chaining large language model prompts through visual
  programming.
\newblock In S.~D.~J. Barbosa, C.~Lampe, C.~Appert, and D.~A. Shamma, eds.,
  {\em {CHI} '22: {CHI} Conference on Human Factors in Computing Systems, New
  Orleans, LA, USA, 29 April 2022 - 5 May 2022, Extended Abstracts}, pp.
  359:1--359:10. {ACM}, 2022. \href{https://doi.org/10.1145/3491101.3519729}
{doi: {{%
10\hspace{.1pt}\discretionary{.}{%
}{.}\hspace{.4pt}1145\discretionary{/}{%
}{/}3491101\hspace{.1pt}\discretionary{.}{%
}{.}\hspace{.4pt}3519729}}}


\bibitem{gpt4v3d}
T.~Wu, G.~Yang, Z.~Li, K.~Zhang, Z.~Liu, L.~J. Guibas, D.~Lin, and
  G.~Wetzstein.
\newblock Gpt-4v(ision) is a human-aligned evaluator for text-to-3d generation.
\newblock In {\em {IEEE/CVF} Conference on Computer Vision and Pattern
  Recognition, {CVPR} 2024, Seattle, WA, USA, June 16-22, 2024}, pp.
  22227--22238. {IEEE}, 2024.
  \href{https://doi.org/10.1109/CVPR52733.2024.02098}
{doi: {{%
10\hspace{.1pt}\discretionary{.}{%
}{.}\hspace{.4pt}1109\discretionary{/}{%
}{/}CVPR52733\hspace{.1pt}\discretionary{.}{%
}{.}\hspace{.4pt}2024\hspace{.1pt}\discretionary{.}{%
}{.}\hspace{.4pt}02098}}}


\bibitem{xia2022interactive}
J.~Xia, L.~Huang, W.~Lin, X.~Zhao, J.~Wu, Y.~Chen, Y.~Zhao, and W.~Chen.
\newblock Interactive visual cluster analysis by contrastive dimensionality
  reduction.
\newblock {\em {IEEE} Trans. Vis. Comput. Graph.}, 29(1):734--744, 2023.
  \href{https://doi.org/10.1109/TVCG.2022.3209423}
{doi: {{%
10\hspace{.1pt}\discretionary{.}{%
}{.}\hspace{.4pt}1109\discretionary{/}{%
}{/}TVCG\hspace{.1pt}\discretionary{.}{%
}{.}\hspace{.4pt}2022\hspace{.1pt}\discretionary{.}{%
}{.}\hspace{.4pt}3209423}}}


\bibitem{yang2021explaining}
L.~Yang, C.~Xiong, J.~K. Wong, A.~Wu, and H.~Qu.
\newblock Explaining with examples: Lessons learned from crowdsourced
  introductory description of information visualizations.
\newblock {\em {IEEE} Trans. Vis. Comput. Graph.}, 29(3):1638--1650, 2023.
  \href{https://doi.org/10.1109/TVCG.2021.3128157}
{doi: {{%
10\hspace{.1pt}\discretionary{.}{%
}{.}\hspace{.4pt}1109\discretionary{/}{%
}{/}TVCG\hspace{.1pt}\discretionary{.}{%
}{.}\hspace{.4pt}2021\hspace{.1pt}\discretionary{.}{%
}{.}\hspace{.4pt}3128157}}}


\bibitem{zeng2022gesturelens}
H.~Zeng, X.~Wang, Y.~Wang, A.~Wu, T.-C. Pong, and H.~Qu.
\newblock Gesturelens: Visual analysis of gestures in presentation videos.
\newblock {\em IEEE Transactions on Visualization and Computer Graphics}, 2022.
  \href{https://doi.org/10.1109/TVCG.2022.3169175}
{doi: {{%
10\hspace{.1pt}\discretionary{.}{%
}{.}\hspace{.4pt}1109\discretionary{/}{%
}{/}TVCG\hspace{.1pt}\discretionary{.}{%
}{.}\hspace{.4pt}2022\hspace{.1pt}\discretionary{.}{%
}{.}\hspace{.4pt}3169175}}}


\bibitem{zeng2019emoco}
H.~Zeng, X.~Wang, A.~Wu, Y.~Wang, Q.~Li, A.~Endert, and H.~Qu.
\newblock Emoco: Visual analysis of emotion coherence in presentation videos.
\newblock {\em IEEE Transactions on Visualization and Computer Graphics},
  26(1):927--937, 2019. \href{https://doi.org/10.1109/TVCG.2019.2934656}
{doi: {{%
10\hspace{.1pt}\discretionary{.}{%
}{.}\hspace{.4pt}1109\discretionary{/}{%
}{/}TVCG\hspace{.1pt}\discretionary{.}{%
}{.}\hspace{.4pt}2019\hspace{.1pt}\discretionary{.}{%
}{.}\hspace{.4pt}2934656}}}


\bibitem{vahdat2022lion}
X.~Zeng, A.~Vahdat, F.~Williams, Z.~Gojcic, O.~Litany, S.~Fidler, and K.~Kreis.
\newblock {LION:} latent point diffusion models for 3d shape generation.
\newblock In S.~Koyejo, S.~Mohamed, A.~Agarwal, D.~Belgrave, K.~Cho, and A.~Oh,
  eds., {\em Advances in Neural Information Processing Systems 35: Annual
  Conference on Neural Information Processing Systems 2022, NeurIPS 2022, New
  Orleans, LA, USA, November 28 - December 9, 2022}, 2022.

\bibitem{zhang2024gaussiancube}
B.~Zhang, Y.~Cheng, J.~Yang, C.~Wang, F.~Zhao, Y.~Tang, D.~Chen, and B.~Guo.
\newblock Gaussiancube: Structuring gaussian splatting using optimal transport
  for 3d generative modeling.
\newblock {\em CoRR}, abs/2403.19655, 2024.
  \href{https://doi.org/10.48550/ARXIV.2403.19655}
{doi: {{%
10\hspace{.1pt}\discretionary{.}{%
}{.}\hspace{.4pt}48550\discretionary{/}{%
}{/}ARXIV\hspace{.1pt}\discretionary{.}{%
}{.}\hspace{.4pt}2403\hspace{.1pt}\discretionary{.}{%
}{.}\hspace{.4pt}19655}}}


\bibitem{zhao2024michelangelo}
Z.~Zhao, W.~Liu, X.~Chen, X.~Zeng, R.~Wang, P.~Cheng, B.~Fu, T.~Chen, G.~Yu,
  and S.~Gao.
\newblock Michelangelo: Conditional 3d shape generation based on
  shape-image-text aligned latent representation.
\newblock In A.~Oh, T.~Naumann, A.~Globerson, K.~Saenko, M.~Hardt, and
  S.~Levine, eds., {\em Advances in Neural Information Processing Systems 36:
  Annual Conference on Neural Information Processing Systems 2023, NeurIPS
  2023, New Orleans, LA, USA, December 10 - 16, 2023}, 2023.

\bibitem{zhu2021visualizing}
H.~Zhu, M.~Zhu, Y.~Feng, D.~Cai, Y.~Hu, S.~Wu, X.~Wu, and W.~Chen.
\newblock Visualizing large-scale high-dimensional data via hierarchical
  embedding of knn graphs.
\newblock {\em Visual Informatics}, 5(2):51--59, 2021.
  \href{https://doi.org/10.1016/j.visinf.2021.06.002}
{doi: {{%
10\hspace{.1pt}\discretionary{.}{%
}{.}\hspace{.4pt}1016\discretionary{/}{%
}{/}j\hspace{.1pt}\discretionary{.}{%
}{.}\hspace{.4pt}visinf\hspace{.1pt}\discretionary{.}{%
}{.}\hspace{.4pt}2021\hspace{.1pt}\discretionary{.}{%
}{.}\hspace{.4pt}06\hspace{.1pt}\discretionary{.}{%
}{.}\hspace{.4pt}002}}}


\end{thebibliography}

\appendix 

\end{document}